\documentclass[aps,twocolumn,prd,showpacs,nofootinbib]{revtex4}
\usepackage{amsmath}
\usepackage{graphicx}
\usepackage{subfigure}
\usepackage{dcolumn}
\usepackage{bm}
\usepackage{amssymb}
\usepackage{latexsym}
\usepackage{psfrag}

\def\setC{\mathbb{C}}

\def\setR{\mathbb{R}}

\newcommand{\ie}{\textsl{i.e.~}}

\newcommand{\etal}{\textsl{et al.~}}

\newcommand{\mP}{m_{_{\mathrm Pl}}}
\newcommand{\GReCO}{${\cal G}\setR\varepsilon\setC{\cal O}$}

\def\spose#1{\hbox to 0pt{#1\hss}}
\def\lta{\mathrel{\spose{\lower 3pt\hbox{$\mathchar"218$}}
     \raise 2.0pt\hbox{$\mathchar"13C$}}}
\def\gta{\mathrel{\spose{\lower 3pt\hbox{$\mathchar"218$}}
     \raise 2.0pt\hbox{$\mathchar"13E$}}}

\newcommand{\dphi}{\delta\varphi}
\newcommand{\mean}[1]{\left\langle #1 \right\rangle}
\newcommand{\corr}[1]{\left\langle \varphi^{#1}(t) \right\rangle}

\newcommand{\de}[2]{\kern - #1 em \mathrm{d} #2}
\newcommand{\ini}{\mathrm{in}}
\newcommand{\cl}{\mathrm{cl}}
\newcommand{\tin}{{t_\mathrm{in}}}
\newcommand{\phiin}{{\varphi_\mathrm{in}}}

\begin{document}

\title{Stochastic Quintessence}

\author{J\'er\^ome Martin} 
\email{jmartin@iap.fr}
\affiliation{Institut d'Astrophysique de
Paris, \GReCO, FRE 2435-CNRS, 98bis boulevard Arago, 75014 Paris,
France}

\author{Marcello Musso}
\email{marcello.musso@pv.infn.it}
\affiliation{Univert\`a di Pavia \& INFN - Sezione di Pavia, 
via U. Bassi 6, I-27100 Pavia, Italy \\
Institut d'Astrophysique de
Paris, \GReCO, FRE 2435-CNRS, 98bis boulevard Arago, 75014 Paris, France}

\date{\today}

\begin{abstract}
The behavior of the quintessence field is studied during inflation. In
order to have a satisfactory model of dark energy, the quintessence
field value today should be as insensible to the initial conditions as
possible. Usually, only the dependence on the initial conditions
specified at the end of inflation or, equivalently, at the beginning
of the radiation dominated era, is considered. Provided the
quintessence field is initially within a large but, crucially, finite
interval, its present value becomes independent of the initial value
it started from. The question as to whether inflation naturally drives
the quintessence field to the above-mentioned interval is
addressed. Since the quantum effects turn out to be important, the
formalism of stochastic inflation is used in order to calculate the
evolution of the quintessence field. Moreover, the quantum effects
originating from the inflaton field are also taken into account and
are proved to be sub-dominant in most cases. Finally, the requirement
that the quintessence field is on tracks today is shown to imply quite
tight constraints on the initial values of the quintessence and
inflaton fields at the beginning of inflation. In particular, the
initial value of the inflaton field cannot be too large which
indicates that the quintessential scenario seems to be compatible with
inflation only if the total number of e-folds is quite small. 
\end{abstract}

\maketitle

\section{Introduction}

The observations suggesting that our Universe is presently undergoing
a phase of accelerated expansion have recently accumulated~\cite{SNIa,
wmap, sdss,ISW}. If really confirmed, this discovery is certainly a
breakthrough for cosmology but, at the same time, represents a big
challenge since finding a convincing explanation for such a phenomenon
is clearly a difficult task.

\par

{}From a theoretical point of view, the presence of a non-vanishing
cosmological constant whose energy density would be of the order of
the critical energy density today seems to be the most natural
solution. In addition, the currently available data on the equation of
state are, so far, compatible with this assumption. But it is
well-known that the theoretical preferred value of the cosmological
constant corresponds to an energy density much larger than the
critical energy density and there exists, at the moment, no convincing
arguments which would explain this difference~\cite{Wein}.

\par

This situation has led the physicists to seek for alternatives. Among
the solutions proposed, the quintessence scenario has recently
attracted a lot of attention~\cite{RP, quint,
ZWP,PB,BM1,BM2,BMR1,BMR2,highphysquint}. It consists in postulating
that the acceleration of the expansion is caused by a scalar field,
the quintessence field $Q$, evolving in a potential the typical shape
of which is given by $W(Q)=M^{4+\alpha }Q^{-\alpha }$, where $M$ is an
energy scale and $\alpha >0$ a free parameter~\cite{RP}. The main
advantage of this scenario is that the coincidence problem can be
solved because the equations of motion possess a solution which is an
attractor. Therefore, the present evolution of the quintessence field
is independent from the choice of the initial conditions. Moreover,
when the field is on tracks, $Q$ is typically of the order of the
Planck mass and hence, for not too small values of $\alpha $, the
scale $M$ can be large. As a result, the fine-tuning is less severe
than in other scenarios because it is possible to explain the presence
of a very small scale (the vacuum energy density today) by means of a
theory which, on the contrary, is characterized by a large scale
$M$. This is due to the inverse power-law shape of the potential and
is reminiscent of the ``see-saw'' mechanism in particle physics. This
has also the advantage that model building can be considered in the
realm of high energy
physics~\cite{PB,BM1,BM2,BMR1,BMR2,highphysquint}.

\par

So far most of the studies have been devoted to understanding how the
quintessence field evolves from the beginning of the radiation era
until now. Another important (related) question, in view of its
observational implications, has been to estimate the value of the
equation of state today. In this paper, we address a new question,
namely that of the behavior of $Q$ during (chaotic) inflation
assuming, for simplicity, that the quintessence field and the inflaton
are not coupled. This is an important problem since it is crucial to
check that $Q$ is, at the end of inflation (or at the beginning of the
radiation dominated epoch), in the range of values which are such that
the field is on tracks today. 

\par

However, the problem does not only boil down to solving the
Klein-Gordon equation in an inflationary background. Indeed, in
Ref.~\cite{ML}, it has been suggested that the quantum effects could
play an important role. In this case, the techniques of stochastic
inflation~\cite{fokkerplanck,stocha} can be used to describe the
evolution of the quintessence field. This was done for the first time
in Ref.~\cite{ML}. The method utilized in that article was to solve
the Fokker-Planck equation in order to follow the evolution of the
probability distribution of the quintessence field. It was then shown
that, typically, the attractor is joined at relatively small
redshifts.

\par

In the present paper, we consider the above-mentioned question again
but from a different perspective. One of our main purposes is to
calculate the probability distribution function of the quintessence
field at the end of inflation (or at the beginning of the radiation
dominated era) in order to estimate whether it is likely that the
value of $Q$ is such that the attractor is joined today. Moreover,
requiring that the corresponding probability is significant can be
used to constrain the space of the initial conditions, \ie the initial
value of the inflaton (or, equivalently, the total number of e-folds)
and quintessence fields. In addition, we demonstrate that this also
puts constraints on the power index $\alpha $ characterizing the shape
of the quintessence potential, namely small values of $\alpha $ are
disfavored.

\par

Another goal of the present work is to include the inflaton
fluctuations, to study under which circumstances their effect can be
important and, when it is the case (and when it is possible), to
calculate the corresponding correction to the behavior of the
quintessence field. Indeed, in Ref.~\cite{ML}, since the inflaton was
treated as a classical field, the quantum effects were only sourced by
the quintessence noise. However, the inflaton itself is also
influenced by the quantum effects and, therefore, a priori the
inflaton noise also affects the evolution of the quintessence
field. In fact, the variance of the quintessence field can be written
as
\begin{equation}
\sigma ^2(t)=\frac{1}{4\pi ^2}\int _{\tin}^t
H^3\left[\varphi \left(\tau \right)\right]{\rm d}\tau  \, ,
\end{equation}
where $H$ is the Hubble parameter. Roughly speaking, taking into
account the inflaton noise amounts to put the coarse-grained inflaton
in the above equation rather that its classical counterpart. If the
expression calculated in this way differs significantly from the
expression obtained by inserting the classical inflaton, then the
inflaton noise plays indeed a non negligible role.

\par

Another difference from Ref.~\cite{ML} is that we directly solve the
Langevin equation rather than the Fokker-Planck equation. Obviously,
this is only a technical difference since the two approaches are
equivalent. In a first time, the Langevin equation is solved by means
of a perturbative expansion. In this regime, we show that the
influence of the inflaton noise is always negligible. Then, in a
second time, we try to solve the Langevin equation in the
non-perturbative regime (for the quintessence field) by modeling the
effect of the classical force with a wall.

\par

This article is organized as follows. In the next section, we quickly
review the basic principles and equations of the stochastic
approach. Then, in Sec.~III, we present the perturbative method used
to solve the Langevin equation. We apply this method to inflation,
compare the results obtained with those already known in the
literature and demonstrate that they are equivalent. In Sec.~IV, we
apply the perturbative approach to the Langevin equation for the
quintessence field and explicitly calculate the quintessential quantum
effects. In this regime, we show that the contribution coming from the
inflaton noise is negligible. Then, we present a model with a
reflecting wall which allows us to explore a region where the
perturbative approach breaks down. We study the constraints on the
initial conditions of the inflaton and quintessence fields that exist
in order for the coincidence problem to still be solved. We prove
that these constraints are quite stringent. Finally, in Sec.~V, we
discuss the results obtained in this article and present our
conclusions.


\section{Basic equations}

In the Friedman-Lema\^{\i}tre-Robertson-Walker (FLRW) Universe, the
metric of which can be written as ${\rm d}s^2=-{\rm d}t^2+a^2(t)
\delta _{ij}{\rm d}x^{i}{\rm d}x^j$ (we assume flat space-like
sections), the evolution of a scalar field $\phi (t,{\bf x})$ is
described by the Klein-Gordon equation
\begin{equation}
\label{KG}
  \ddot\phi+3H\dot\phi-\frac{\nabla^2\phi}{a^2}
+\frac{{\rm d}V(\phi)}{{\rm d}\phi }=0\, ,
\end{equation}
where a dot denotes the derivation with respect to the cosmic time
$t$. 

\par

In the stochastic formalism~\cite{stocha}, one is interested in the
dynamics of a ``coarse-grained'' field $\varphi (t,{\bf x})$. This
coarse-grained field is defined to be the average of the ordinary
field $\phi$ over a physical volume whose size is somewhat larger than
the Hubble radius $H^{-1}\equiv a/\dot{a}$. Therefore, $\varphi(t,{\bf
x})$ basically contains the long-wavelength Fourier modes (with
wavenumber $k<aH$) only. Technically, one writes
\begin{widetext}
\begin{equation}
\label{corse}
\phi (t,{\bf x})=\varphi (t,{\bf x})+\frac{1}{(2\pi )^{3/2}}
\int {\rm d}{\bf k}W\left(k-\sigma aH\right)\left[c_{\bf k}\mu _k(t)
{\rm e}^{i{\bf k}\cdot {\bf x}}+c_{\bf k}^{\dagger }\mu _k^*(t)
{\rm e}^{-i{\bf k}\cdot {\bf x}}\right]\, ,
\end{equation}
\end{widetext}
where $W(z)$ is the so-called window function. In the case of a white
noise, the window function is the step function. In a more realistic
situation, the window function should be taken as a smoothed version
of the step function~\cite{colored}. This corresponds to the case of a
colored noise and the problem is generally technically more
complicated in this situation. In this article, for simplicity, we
restrict ourselves to the case of a white noise. It should also be
noticed that, in Eq.~(\ref{corse}), the mode functions $\mu _k(t)$
are, by definition, the free mode functions, \ie obey the equation
$\ddot{\mu }_k+3H\dot{\mu }_k+(k^2/a^2)\mu _k=0$. Finally, $\sigma $
is a parameter smaller than $1$, introduced in order to allow some
level of arbitrariness in the choice of the smoothing scale.

\par

The evolution of the coarse-grained field is still described by the
Klein-Gordon equation \eqref{KG} but a suitable random noise field
$\xi (t)$, acting as a classical stochastic source term, should be
added to the right hand side in order to mimic the quantum
fluctuations. In the slow-roll approximation $\ddot\varphi$ is
negligible compared to $3H\dot\varphi$ and, since we are dealing with
super-Hubble scales, the gradient term can also be dropped. The
coarse-grained field is thus governed by a first order Langevin-like
differential equation which can be put in the form
\begin{equation}
\label{Langevin}
\frac{{\rm d}\varphi}{{\rm d}t}+\frac{1}{3H}\frac{{\rm d}V}{{\rm
  d}\varphi}= \frac{H^{3/2}}{2\pi}\xi(t)\, ,
\end{equation}
where the noise field $\xi$ is defined in such a
way that its correlation function simply reads
\begin{equation}
\label{noisecorr}
  \left\langle\xi(t)\xi(t')\right\rangle=\delta(t-t')\, .
\end{equation}
where $\delta(z)$ is the Dirac function. The normalization of the
correlation function is chosen in order to reproduce, for a free
field, the ordinary de Sitter result $\mean{\varphi^2}=H^3t/(4\pi
^2)$.

\par

At this point, two situations are possible, leading to very different
technical problems. The first possibility corresponds to the case
where the scalar field is a test field in a fixed background. This
means that the factors $H$ which appear into the Langevin equation (at
the denominator in the second term and at the numerator in the third
term) must be considered as functions of time but not as functions of
the coarse-grained field. This is obviously an important
simplification and, in this case, the noise is said to be
non-multiplicative. In such a situation, the derivation of the
Langevin equation is unambiguous and on a firm basis. In order to see
how the formalism works, let us quickly consider the case where $H$ is
constant in time (\ie de Sitter background) and $V(\phi)=m^2\phi^2/2$.
Then, the solution of the Langevin equation can be found explicitly
yielding
\begin{eqnarray}
  \varphi(t) &=& {\rm e}^{-m^2\left(t-t_{\rm in}\right)/(3H)}
\nonumber \\ & \times & \left[\varphi_{\rm in} +
\frac{H^{3/2}}{2\pi}\int^t_{t_{\rm in}}{\rm d}\tau \, {\rm
e}^{m^2\left(\tau -t_{\rm in}\right)/(3H)}\xi\left(\tau
\right)\right]\, ,
\end{eqnarray}
where $\varphi _{\rm in}$ is the initial value of the field. Then, we
can easily deduce the two-point function and we obtain
\begin{equation}
  \corr{2}=\frac{3H^4}{8\pi^2 m^2} + 
  \left(\varphi_{\rm in}^2-\frac{3H^4}{8\pi^2 m^2}\right)
  {\rm e}^{-2m^2\left(t-t_{\rm in}\right)/(3H)}\, .
\end{equation}
After a transitory regime, one sees that the two-point correlation
function goes to $\corr{2}\rightarrow 3H^4/(8\pi^2 m^2)$. This
well-known result has already been obtained, for instance in
Ref.~\cite{SY} by solving the Fokker-Planck equation. Let us notice
that we could have also found the solution and computed the
correlation function in the case of a colored noise.

\par

The second situation corresponds to the situation where one takes into
account the back-reaction of the coarse-grained field on the geometry.
Technically, this means that the Hubble parameter in
Eq.~(\ref{Langevin}) becomes a function of the coarse-grained field
itself. In other words, the noise becomes multiplicative. In this
case, it is necessary to have one more equation and one naturally
assumes that the Friedman equation (in the slow-roll approximation)
holds for the coarse-grained quantities, namely
\begin{equation}
\label{einstein}
  H^2(\varphi)\simeq \frac{8\pi}{3\mP^2}V(\varphi)
  \equiv\frac{\kappa}{3}V(\varphi)\, .
\end{equation}
This case if of course the most interesting since it corresponds to
the case of inflation. The coarse-grained field becomes the
coarse-grained inflaton which drives the evolution of the background.

\par

Unfortunately, as is well-known in the case of a multiplicative noise,
the derivation of the Langevin equation becomes also more problematic,
see for instance Refs.~\cite{pbs}. Roughly speaking, this is due to
the following. When we promote the field $\varphi $ to a stochastic
quantity, there is some arbitrariness in defining the term $H^{3/2}\xi
$ in the Langevin equation. Indeed, the quantity $H^{3/2}$ originates
from two contributions. The first one comes from the term $H$ present
in the damping term of the Klein-Gordon equation which, according to
the rule outlined above, should be promoted to a stochastic
quantity. The second one comes from the normalization of the noise
correlation function which is an ordinary function. Therefore, the
problem arises because one could choose to promote a different power
of the Hubble constant to a stochastic quantity, say $H^{3/2-x}$ and
put the remaining term, $H^x$, into the normalization of the noise. In
this case, the Langevin equation would lead to different results. In
the present paper, we consider $H^{3/2}$ as a stochastic
quantity. Finally, we notice that there is also the ambiguity in the
choice of the calculus. Here, we work with the Stratonovitch calculus.

\section{Inflation}

\subsection{Classical Evolution}

Having specified what our basic set up is, we now turn to the
application of this formalism to inflation. For simplicity, in the
following, we restrict ourselves to single-field ``chaotic'' inflation
models~\cite{linde}. It turns out useful to parameterize the potential
in term of the dimensionless scalar field $\varphi/\mP$. Explicitly,
we take (with $n\geq 2$)
\begin{equation}
\label{pot}
  V(\varphi)=  V_0\left(\frac{\varphi}{\mP}\right)^n \, .
\end{equation}
The Cosmic Microwave Background Radiation (CMBR) anisotropy
observations constrain the value of $V_0$. For small $\ell $, the
multipole moments are given by
\begin{equation}
C_{\ell }=\frac{2H^2}{25\epsilon \mP^2}\frac{1}{\ell (\ell +1)}\, 
\end{equation}
and what has been actually measured by the COsmic Background Explorer
(COBE)~\cite{cobe} and the Wilkinson Microwave Anisotropy Probe
(WMAP)~\cite{wmap} satellites is $Q_{\rm rms-PS}^2/T^2=5C_2/(4\pi
)\simeq 18\times 10^{-6}/(2.7)\simeq 6\times 10^{-6}$.  The quantity
$\epsilon $ is the first slow-roll parameter~\cite{sr} and for chaotic
models, it reads $\epsilon \simeq n/(4N_*+n)$ where $N_*\simeq 60$ is
the number of e-folds between Hubble radius exit and the end of
inflation. Putting everything together, we find that $V_0$ is given by
\begin{equation}
\frac{V_0}{\mP^4}\simeq
\frac{90}{(4N_*+n)^{n/2+1}}\left(\frac{16}{n}\right)^{n/2}
\frac{Q_{\rm rms-PS}^2}{T^2}\, .
\end{equation}
{}From an observational point of view, all the models such that $n>5$
are now excluded by the WMAP data, the quartic case being on the
border line, see Ref.~\cite{LL}.

\par

For the simple potentials considered here, the slow-roll equations can
be integrated exactly. For this purpose, it is convenient to express
everything in terms of the total number of e-folds defined by
\begin{equation}
N\equiv \ln \left(\frac{a}{a_{\rm in}}\right)\, ,
\end{equation}
such that, initially, one has $N=0$. Then, the classical field, \ie
the solution to the slow-roll equations of motion without the noise,
reads
\begin{equation}
\label{solclass}
\frac{\varphi_{\rm cl}}{\mP}=\sqrt{\left(\frac{\varphi _{\rm
in}}{\mP}\right)^2 -\frac{n}{4\pi }N}\, ,
\end{equation}
where $\varphi_{\rm cl}(N=0)=\phiin$. The model remains under control
only if the energy density is below the Planck energy density. This
amounts to the following constraint on the initial conditions
$\phiin/\mP \lta (\mP^4/V_0)^{1/n}$. Inflation stops when the
slow-roll parameter $\epsilon $ is equal to unity corresponding to
$\varphi _{\rm end}/\mP=n/(4\sqrt{\pi })$. As a consequence, one can
easily check that the argument of the square root in
Eq.~(\ref{solclass}) remains always positive. Finally, the total
number of e-folds during inflation is simply given by $N_{_{\rm
T}}=4\pi (\phiin/\mP)^2/n-n/4$. This number can be huge if the initial
energy density of the inflaton field is close to the Planck energy
density.

\subsection{Perturbative Solutions}

In general, the Langevin equation cannot be solved analytically even
for the simple potentials given by Eq.~(\ref{pot}). Therefore, we use
perturbative techniques.  We consider the coarse-grained field
$\varphi$ as a perturbation of the solution $\varphi _{\rm cl}$ of the
classical equation, \ie we write
\begin{equation}
\varphi(t)=\varphi_\cl(t)+\dphi_1(t)+\dphi_2(t)+\cdots \, ,
\end{equation}
where the term $\dphi_i(t)$ depends on the noise at the power
$i$. Clearly, this expansion is valid as long as
$\dphi_2<\dphi_1<\varphi_\cl$. Expanding up to second order in the
equation of motion, we get two linear differential equations for
$\dphi_1$ and $\dphi_2$, namely
\begin{equation}
  \frac{{\rm d}\dphi_1}{{\rm d}t}+\frac{2}{\kappa}H''(\varphi _{\rm
  cl}) \dphi_1 = \frac{H^{3/2}(\varphi _{\rm cl})}{2\pi}\xi(t)
\end{equation}
and
\begin{eqnarray}
 & &  \frac{{\rm d}\dphi_2}{{\rm d}t}+\frac{2}{\kappa}H''(\varphi _{\rm
  cl}) \dphi_2 = -\frac{H'''(\varphi _{\rm cl})}{\kappa}\dphi_1^2 
\nonumber \\  
& & +
\frac{3}{4\pi}H^{1/2}(\varphi _{\rm cl})H'(\varphi _{\rm
  cl})\dphi_1\xi(t)\, ,
\end{eqnarray}
where a prime denotes a derivative with respect to the field. These
equations can be solved by varying the integration constant.
\begin{figure*}[t]
  \includegraphics[width=.90\textwidth,height=.50\textwidth]{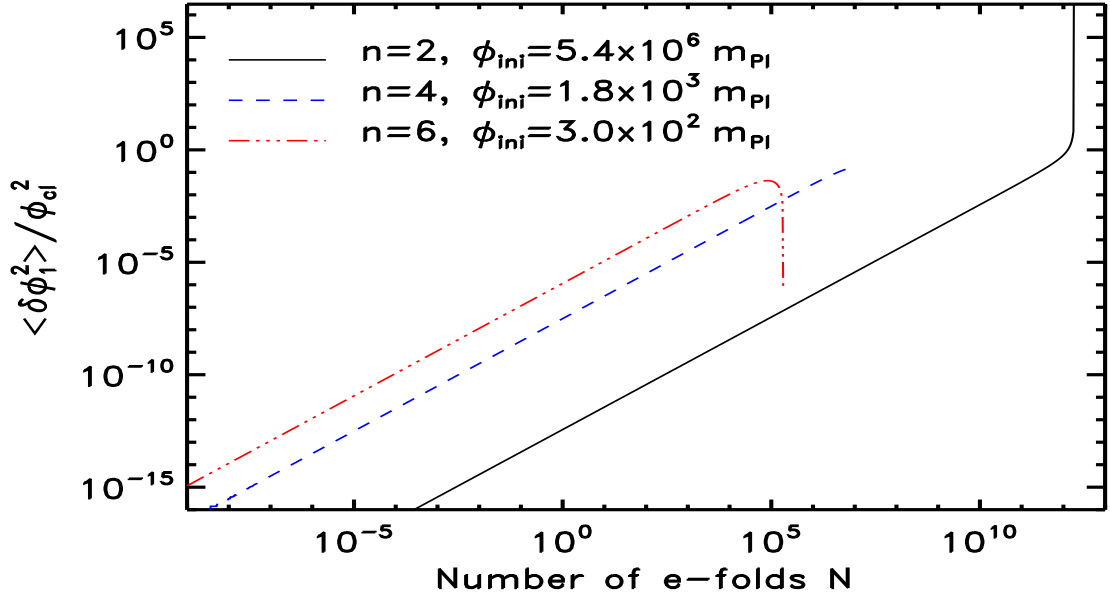} 
\caption{\label{var_infl_fig}
Evolution of $\mean{\delta \varphi _1^2}/\varphi _\cl^2$
versus the number of e-folds during inflation for three different
chaotic potentials characterized by $n=2,4,6$. The initial values of
the inflaton field corresponds to an initial energy density of
$V_\ini=0.5\mP^4$ and implies that the total number of e-folds during
inflation is not the same for the three models under
considerations. This is also the reason why the curves do not stop at
the same place. For the $n=2$ model, $\mean{\delta \varphi _1^2}$
always increases and, at some point, the perturbative scheme used to
calculate it breaks down. For $n=4$, $\mean{\delta \varphi
_1^2}/\varphi _\cl^2$ tends to a constant that is difficult to
visualize because of the e-folds logarithmic scale. For $n \ge 6$,
$\mean{\delta \varphi _1^2}$ possesses a maximum during inflation and
then decreases.}
\end{figure*}
Let us first consider the equation for $\dphi_1$. If the initial
conditions are such that $\dphi_1(t=\tin)=0$, then the solution reads
\begin{equation}
\label{dph1inter}
  \dphi_1(t)=\int_\tin^t\de{.5}{\tau}\frac{H^{3/2}(\varphi _{\rm
  cl})}{2\pi}
  \exp\!\left[-\int_\tau^t\de{.5}{\sigma}\frac{2H''(\varphi _{\rm
  cl})}{\kappa}\right] \xi(\tau)\, .
\end{equation}
This expression can be further simplified. If we use the classical
equation of motion, then one can write the exponential term as
\begin{equation}
  \exp\!\left(-\int_\tau^t\de{.5}{\sigma}\frac{2H''}{\kappa}\right)=
  \exp\!\left[\int_{\varphi _{\rm cl}(\tau)}^{\varphi_{\rm cl}(t)}
\de{1}{\varphi}
  \frac{H''}{H'}\right]=
  \frac{H'\left[\varphi(t)\right]}{H'\left[\varphi(\tau)\right]}\, .
\end{equation}
Inserting the above expression into Eq.~(\ref{dph1inter}), one finally
arrives at
\begin{equation}
\label{soldphi1}
  \dphi_1(t)=\frac{H'\left[\varphi _{\rm
  cl}(t)\right]}{2\pi}\int_\tin^t\de{.5}{\tau}
  \frac{H^{3/2}\left[\varphi _{\rm cl}(\tau )\right]}{H'\left[\varphi
  _{\rm cl}(\tau )\right]}\xi(\tau)\, .
\end{equation}
We are now in a position where the various correlation functions can
be calculated exactly. Since $\dphi_1$ is linear in the noise $\xi$,
the mean value obviously vanishes
\begin{equation}
  \mean{\dphi_1}=0\, .
\end{equation}
Let us now evaluate the two-point correlation function calculated at
the same time, \ie the variance. Making use of Eq.~(\ref{noisecorr}),
one obtains
\begin{equation}
  \mean{\dphi_1^2} = 
  \frac{\kappa}{2}\left(\frac{H'}{2\pi}\right)^2 
  \!\!\int^{\varphi_\ini}_{\varphi _{\rm cl}}\de{.7}{\varphi}
  \left(\frac{H}{H'}\right)^3\, .
\end{equation}
So far, the expressions presented above are general and do not rely on
a particular shape of the inflaton potential. If we now specify the
calculation to the chaotic potential given by Eq.~(\ref{pot}), then
the variance takes the form
\begin{equation}
\label{varinflaton}
  \frac{\mean{\dphi_1^2}}{\mP^2}=-\frac{4}{3n}\frac{V(\varphi _\cl)}{\mP^4}
  \left(\frac{\varphi _{\rm cl}}{\mP}\right)^{-2} 
\left[\left(\frac{\varphi _{\rm cl}
  }{\mP}\right)^4-\left(\frac{\varphi _{\rm
  in}}{\mP}\right)^4\right]\, .
\end{equation}
Since $\varphi _{\rm cl}$ is always smaller than $\phiin$ (because the
field rolls down its potential), the above quantity is increasing with
time and always positive as required.

\par

We now turn to the equation of motion for the second order
perturbation $\dphi_2$. It can be solved by following exactly the
steps that were described before. Then, the solution can be written as
\begin{eqnarray}
  \dphi_2(t) &=& -\frac{H'}{\kappa }\int _\tin^t \de{.5}\tau
\frac{H'''}{H'}\delta \varphi _1^2(\tau ) \nonumber \\ & &
+\frac{3H'}{4\pi}\int_\tin^t\de{.5}{\tau}
H^{1/2}\dphi_1(\tau)\xi(\tau) \, .
\end{eqnarray}
As expected the second order perturbation is quadratic in the
noise. One can now easily evaluate the mean value of $\dphi_2(t)$,
taking into account a factor $1/2$ which originates from the fact that
the Dirac $\delta$-function appearing in the noise correlation
function is centered on an integration limit. One obtains
\begin{widetext}
\begin{eqnarray}
\frac{\mean{\dphi_2}}{\mP} &=& 
-\frac{n+2}{3n}\frac{V(\varphi _\cl)}{\mP^4}
\left(\frac{\varphi _{\rm cl}}{\mP}\right)^{-n/2-1}
\left[\left(\frac{\varphi _{\rm cl}}{\mP}\right)^{n/2+2}
-\left(\frac{\varphi _{\rm in}}{\mP}\right)^{n/2+2}\right]
\nonumber \\
& & +\frac{n-2}{3n}
\frac{V(\varphi _\cl)}{\mP^4}\left(\frac{\varphi _{\rm
in}}{\mP}\right)^4\left(\frac{\varphi _{\rm cl}}{\mP}\right)^{-n/2-1}
\left[\left(\frac{\varphi _{\rm cl}}{\mP}\right)^{n/2-2}
-\left(\frac{\varphi _{\rm in}}{\mP}\right)^{n/2-2}\right]\, .
\label{meanphi}
\end{eqnarray}
\end{widetext}
\begin{figure*}[t]
  \includegraphics[width=.90\textwidth,height=.50\textwidth]{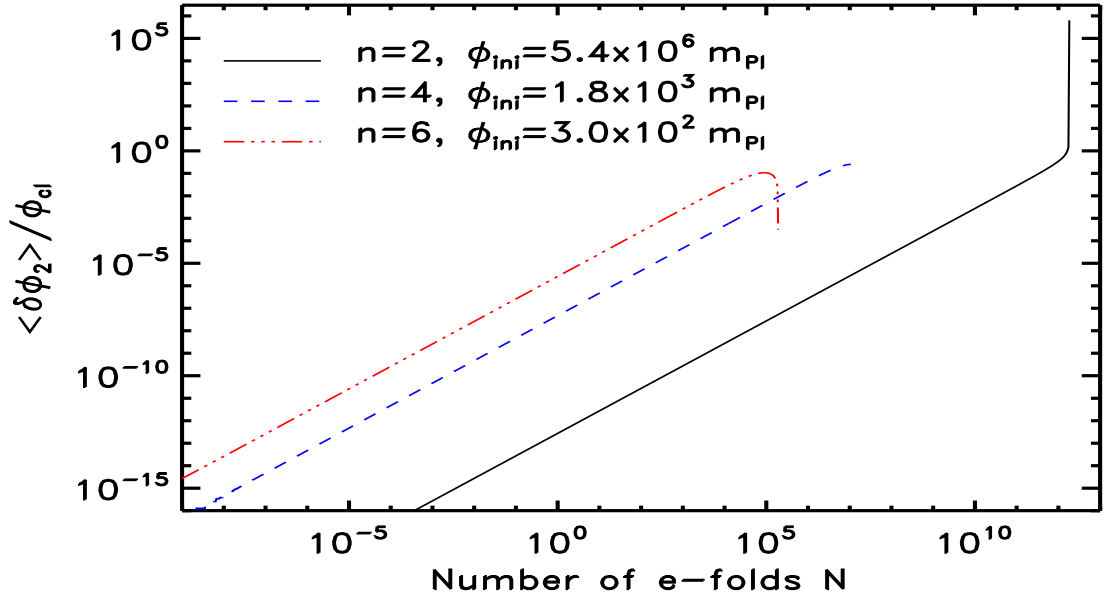}
\caption{\label{mean_infl_fig} Evolution of $\mean{\delta \varphi
_2}/\varphi _\cl$ with the number of e-folds during inflation for
three different potentials. The initial values of the scalar field
always correspond to an initial energy density of
$V_\ini=0.5\mP^4$. As a consequence, the total number of e-folds
during inflation is not the same in these three models and this
explains why the three curves stop at different $N$.  A value of
$\mean{\delta \varphi _2}/\varphi _\cl$ greater than one signals a
breakdown of the perturbative approach. For instance, this is the case
for the $n=2$ model at the end of inflation.}
\end{figure*}
The quantities $\mean{\dphi_1^2}/\varphi _{\rm cl}^2$ and
$\mean{\dphi_2}/\varphi _{\rm cl}$ are represented in
Fig.~\ref{var_infl_fig} and \ref{mean_infl_fig} for different
potentials, \ie for different values of the power index $n$. We find
that for $n>4$ both quantities reach a maximum after an increasing
phase, and then decrease to zero at late times, always remaining
smaller than unity for any choice of the initial conditions (thus
fully validating the perturbative treatment).  Conversely, we see that
their late time behavior is constant for $n=4$ or rapidly increasing
for $n<4$, and appears in both cases to violate the validity of the
perturbative approach for the extreme choice of an initial condition
corresponding to an energy density near the Planck scale.

\subsection{Classicalization}

Let us study the late time behavior of $\mean{\dphi _2}$, as given by
Eq.~(\ref{meanphi}), in more details. In this regime, one has $\varphi
_\cl\ll \varphi _\ini$. Then, it follows that 
\begin{equation}
\frac{\mean{\dphi_2}}{\mP}\propto \left(\frac{\varphi
_\cl}{\mP}\right)^{n/2-1}\, .
\end{equation}
From the above equation, one deduces that we have exact
classicalization for the cases $n=2$ or $n=4$. Indeed, in this case,
we have $\mean{\varphi} \simeq \varphi_\cl+\mean{\dphi_2}\propto
\varphi _\cl$ which is, by definition, what we mean by
classicalization. For $n>4$, the situation is slightly different. In
this case, $\mean{\varphi}$ is not exactly proportional to $\varphi
_\cl$ but the extra contribution $\mean{\dphi_2}$ decays faster than
$\varphi _\cl$ and, hence, becomes negligible as can also be checked
in Fig.~\ref{mean_infl_fig}. In this way, we also recover a classical
behavior.

\par

Let us now see what happens to a trajectory which is far from the mean
value. From Eq.~(\ref{varinflaton}), it is easy to check that, at late
times, we also have $\sqrt{\mean{\dphi_1^2}}/\mP\propto (\varphi
_\cl/\mP) ^{n/2-1}$ (a quantity which gives an estimate of the
amplitude of the probability distribution). It follows than not only
the mean value of the distribution but also its width evolves
according to the classical solution.

\par

Finally, the following remark is in order. The fact that the quantum
effects are negligible does not mean that, at late times, the value of
the quintessence field $\mean{\varphi}$ is the same as the
corresponding classical field value calculated with the same initial
conditions. Since the quantum could have accumulated in the early
phase of evolution causing a deviation from the classical trajectory,
this actually means that the late time value of $\mean{\varphi }$ can
only be viewed as $\varphi _\cl$ but calculated with different initial
conditions.

\subsection{Comparison with Other Methods}

In this section, we compare the results of the previous section with some
known results of the literature.

\par

In Refs.~\cite{hodges,vish}, it has been shown that there is a case
where the Langevin equation can be solved exactly. In order to permit
a more direct comparison with those studies, we write the constant
$V_0$ which appears in the expression of the potential~(\ref{pot}) as
$V_0/\mP^4=3\lambda _n/(8\pi )$, \ie in terms of the coupling constant
$\lambda _n$. Then, the reasoning goes as follows. The Langevin
equation can be written as
\begin{eqnarray}
\label{Langevin2}
& & \frac{{\rm d}}{{\rm d}t}\left(\frac{\varphi
}{\mP}\right)+\frac{n\sqrt{\lambda _n}}{8\pi }m_{_{\rm
Pl}}\left(\frac{\varphi}{\mP}\right)^{n/2-1}
\nonumber \\
& & = \frac{\lambda
_n^{3/4}}{2\pi}m_{_{\rm Pl}}^{1/2}\left(\frac{\varphi
}{\mP}\right)^{3n/4}\xi(t)\, .
\end{eqnarray} 
There is an exact solution if $n=4$ because, in this case, the
Langevin equation takes the form of a Bernoulli equation. The
simplification comes from the fact that the equation can be put under
the form of a linear equation for some power of the field (in the
present context, this power is $-2$ hence the form of the next
equation). Let us also notice that the choice $n=4/3$ would also lead
to a Bernoulli equation but this case looks less interesting since the
power index of the potential is no longer an integer. The Bernoulli
equation can be solved exactly because it may be brought into a linear
form by a change of variable. The solution can be written as
\begin{eqnarray}
\left(\frac{\varphi }{\mP}\right)^{-2}&=&{\rm e}^{\sqrt{\lambda _4}\mP
\left(t-t_{\rm in}\right)/\pi }
\biggl[\left(\frac{\varphi _{\rm in}}{\mP}\right)^{-2}
-\frac{\lambda _4^{3/4}}{\pi }\mP^{1/2}
\nonumber \\
& & \times \int _{t_{\rm in}}^t{\rm d}\tau \, 
{\rm e}^{-\sqrt{\lambda _4}\mP
\left(\tau -t_{\rm in}\right)/\pi }\xi (\tau )\biggr]\, .
\end{eqnarray}
Taking the inverse square root of the general solution one 
then obtains 
\begin{eqnarray}
\label{solphi}
  \varphi(t) = \varphi_\cl(t) \left[1 - \Psi(t)\right]^{-1/2}\, ,
\end{eqnarray}
where $\varphi_\cl(t)$ is the classical solution that can be expressed
explicitly in terms of cosmic time
\begin{equation}
\label{phi4class}
  \varphi_\cl(t)=\varphi_\ini \exp\left[-\frac{\sqrt{\lambda
  _4}\mP}{2\pi }\left(t -t_\ini\right)\right],
\end{equation}
while the quantity $\Psi (t)$ is defined by
\begin{equation}
  \Psi (t)\equiv \frac{\lambda _4^{3/4}}{\pi}\mP^{1/2}
  \int_{t_\ini}^t{\rm d}\tau \left[\frac{\varphi
  _\cl(\tau)}{\mP}\right]^2\xi(\tau )\, .
\end{equation}
This function can be treated as a new dimensionless Gaussian noise
with vanishing mean value and whose variance reads
\begin{align}
\label{varnewnoise}
  \mean{\Psi^2(t)} = \frac{\lambda _4}{2\pi}\left[
  \left(\frac{\varphi_\ini}{\mP}\right)^4-
\left(\frac{\varphi_\cl}{\mP}\right)^4\right]\, .
\end{align}
We are now in a position where the various correlation functions of
the scalar field can be computed. Here, we compute the two-point
correlation function. Using Eq.~(\ref{solphi}), one gets
\begin{equation}
\label{series}
\left\langle \left(\frac{\varphi }{\mP}\right)^2\right\rangle
=\left(\frac{\varphi _{\rm cl}}{\mP}\right)^2 
\sum _{k=0}^{+\infty } \left
\langle \Psi ^k\right \rangle \, .
\end{equation}
Let us notice that we could have also computed any (\ie $n$-point)
correlation function of the field (including of course the mean value)
using the same technique. Since we have to deal with a Gaussian
process, only the even correlation function are non vanishing. In
addition, for a Gaussian process we have
$\mean{\Psi^{2k}}=(2k-1)!!\mean{\Psi^2}^k$. Therefore, we obtain a
series the general term of which can be written as $2^k\Gamma
\left(k+1/2 \right) \mean{\Psi^2}^k/\sqrt{\pi }$, where we have used
Eq.~(8.339.2) of Ref.~\cite{Grad}. At this point two remarks are in
order. Firstly, it is easy to convince oneself that the
series~(\ref{series}) is in fact divergent. The interpretation of this
fact is of course a little bit problematic but, on the other hand, it
is well-known that perfectly well-defined distribution functions can
have no moments. Below, we also give another interpretation of this
fact. Secondly, looking at the expression of the new Gaussian noise
$\Psi $, we see that the series is in fact an expansion in the
coupling constant $\lambda _4$. Since this one is in fact tiny, one
can try to work with the first term of the series only. Then, one
obtains
\begin{eqnarray}
\label{solserieslambda}
\frac{\varphi }{\mP}
&\simeq & \frac{\varphi _{\rm cl}}{\mP}\left[1+\frac12\Psi(t)
+\frac38\Psi ^2(t)+{\cal O}\left(\lambda _4^{9/4}\right)\right]\, .
\end{eqnarray}
This expression is in fact exactly similar to the one obtained
previously for $\varphi _{\rm cl}(t)+\dphi_1(t)+\dphi_2(t)$ and
therefore leads to the same correlation functions as before. It is
easy to show that the expansion in the coupling constant is nothing
but the expansion in the noise used before. Our method is more general
because it is not restricted to the case $n=4$. This is because the
expansion is directly performed in the Langevin equation rather than
in its solution. The drawback of this last method is clearly that it
is first necessary to find a solution of the Langevin equation, which
is not an easy task, before the expansion can be taken. Let us
emphasize again that, even if an exact solution is known, an expansion
is still necessary because only some power of the field is generally
obtained (in the present case $\varphi ^{-2}$) and, in order to find
the expression of the stochastic field itself, one should then compute
the root of the solution.

\par

The same conclusions can be obtained by means of a model where the
random walk of the field is modified by the presence of a reflecting
barrier. This model is explored in the Appendix~\ref{wallinf}. 

\par

Another possibility studied in the literature is the use of the
so-called scaling solutions, see Ref.~\cite{MOL}. Let us quickly review
the method. The first step consists in rendering the stochastic
process non multiplicative. This can be done by means of the
transformation $x=16\times 3^{3/4}\pi^4\int _{-\infty}^{\varphi
}V^{-3/4}(\theta){\rm d}\theta$ which reduces the Langevin equation to
${\rm d}x/{\rm d}t=-{\rm d}\tilde{V}/{\rm d}x+\xi (t)$ where
$\tilde{V}=-3\times (4\pi ^2)^5/V[\varphi (x)]$. Explicitly, one has
$-{\rm d}\tilde{V}/{\rm d}x=g\times(x/\mP)^{(n+4)/(3n-4)}$ where $g$
is a constant which depends on $V_0$ and $n$. The next step is to
consider the time-dependent nonlinear transformation
\begin{equation}
\eta (x,t)=F^{-1}[{\rm e}^{-\gamma (t-\tin)}F(x)]\, ,
\end{equation}
where the function $F$ is defined by, see Ref.~\cite{MOL}
\begin{eqnarray}
F(x) &=& \frac{x_{\rm in}}{\mP}\exp\biggl\{\frac{\gamma (3n-4)}{2g(n-4)}
\biggl[\left(\frac{x}{\mP}\right)^{2(n-4)/(3n-4)}
\nonumber \\
& & -\left(\frac{x_{\rm in}}{\mP}\right)^{2(n-4)/(3n-4)}\biggr]\biggr\}\, ,
\end{eqnarray}
with $\gamma \equiv g(x_{\rm in}/\mP)^{2(4-n)/(3n-4)}$. Then, it is
straightforward to show that the new stochastic process $\eta $ obeys
the following equation
\begin{eqnarray}
\label{Langeta}
\frac{{\rm d}\eta }{{\rm d}t} &=& \biggl[1+2g\frac{n-4}{3n-4}(t-\tin)
\nonumber \\
& & \times
\eta ^{2(4-n)/(3n-4)}\biggr]^{-(n+4)/[2(n-4)]}\xi(t)\, .
\end{eqnarray}
So far, everything is exact. However, the above Langevin equation
cannot be solved exactly. The so-called scaling solutions correspond
to replacing $\eta $ by $\eta _{\rm cl}$ in the right hand side of
Eq.~(\ref{Langeta}). Then, clearly, the Langevin equation can be
integrated in this approximation. Therefore, the scaling solutions are
nothing but the result of an expansion in the noise and, again, are
similar to the solutions found at the beginning of this section. We
notice, as shown in Ref.~\cite{MOL}, that the quantity $\eta _{\rm
cl}$ is in fact a constant. This is consistent with the fact that this
is the zeroth order solution (\ie for which the noise term is simply
neglected) of the above-mentioned expansion for which the Langevin
equation simply reads ${\rm d}\eta /{\rm d}t=0$. Finally, in
Ref.~\cite{MOL}, a saddle point approximation is used in order to
calculate the effective dispersion. The result found is, see Eq.~(28)
of Ref.~\cite{MOL}
\begin{eqnarray}
\mean{\varphi ^2}_{\rm eff} &\propto &H^2(\varphi _{\rm
cl})\sinh\left[\ln H^{-4/n}(\varphi _{\rm cl})\right]
\\
&=& \varphi ^n\left(\varphi ^2+\varphi ^{-2}\right)\, ,
\end{eqnarray}
which is exactly the result obtained in Eq.~(\ref{varinflaton}). This
reinforces the result that the scaling solutions are very similar or
even identical to the solutions exhibited here by means of the
expansion in the noise term. 

\par

In conclusion, the solutions obtained previously for the stochastic
inflaton are explicit and consistent with those already found in the
literature. In the sequel, we use them as a description of the
background in which the quintessence field lives.


\section{Quintessence}

\subsection{Classical Evolution}

In order to explain the accelerated expansion of the universe, one
postulates the presence of the quintessence field $Q$. This field is a
test field during almost all the cosmic evolution and becomes dominant
only recently when it drives the accelerated expansion. As is
well-known for a scalar field, the equation of state $\omega
_Q=p_Q/\rho _Q$ is time-dependent and, crucially, can be negative. The
detailed evolution of $\omega _Q$ clearly depends on the shape of the
quintessence potential $W(Q)$. An interesting choice is the inverse
power law potential (with $\alpha >0$) which was first studied by
Ratra and Peebles in Ref.~\cite{RP}
\begin{equation}
\label{ratra}
  W(Q)=W_0\left(\frac{Q}{\mP}\right)^{-\alpha }.
\end{equation}
During the radiation dominated era, it is possible to find an exact
solution of the corresponding Klein Gordon equation for which
$Q\propto a^{4/(\alpha +2)}$ or $\rho _Q\propto a^{-4\alpha /(\alpha
+2) }$. This is also possible for the matter dominated era for which
one has $Q \propto a^{3/(\alpha +2)}$ or $\rho _Q \propto a^{-3\alpha
/(\alpha +2)}$. The two solutions we have just mentioned can also be
expressed by means of the following equation~\cite{ZWP}
\begin{equation}
\label{attractor}
\frac{{\rm d}^2W(Q)}{{\rm d}Q^2}=\frac{9}{2}\frac{\alpha +1}{\alpha
}(1-\omega _Q^2)H^2\, ,
\end{equation}
this expression being valid both during the radiation and matter
dominated epochs.  We can re-write the parameter $\omega _Q$ as
$\omega _Q=(\alpha \omega _{_{\rm B}}-2)/(\alpha +2)$ where $\omega
_{_{\rm B}}$ is the equation of state of the background, \ie either
$1/3$ or $0$. Since $\rho _Q$ redshifts slower than the background
energy density, the scalar field contribution will eventually become
dominant. From the above equation, it is easy to see that, when the
quintessence field is about to dominate, its value is in fact of the
order of the Planck mass. Then, the value of $W_0$ is constrained by
the fact that the quintessence energy density is almost the critical
energy density today. This gives
\begin{equation}
\frac{W_0}{\mP^4}\simeq \frac{3}{8\pi }\frac{H_0^2}{\mP^2}\, ,
\end{equation}
where $H_0$ is the Hubble parameter today, \ie $H_0\simeq
10^{-61}\mP$. Equipped with this relation, one can also estimate 
the ratio $W_0/V_0$ and one gets
\begin{equation}
\frac{W_0}{V_0}\simeq \frac{(4N_*+n)^{n/2+1}}{240\pi }
\left(\frac{n}{16\pi }\right)^{n/2}\frac{H_0^2}{\mP^2}
\left(\frac{Q_{\rm rms-PS}}{T}\right)^{-2}\, .
\end{equation}
This number is obviously extremely small, $\simeq {\cal
O}(10^{-110})$.

\par

The main property of the solutions described before is that they are
attractors~\cite{RP}. This means that there is no need to fine-tune
the initial conditions and that the solution will be on tracks today
for a large range of initial conditions. Let us be more precise about
this particular point. Usually, one fixes the initial conditions at
the end of inflation, at a redshift of $z=10^{28}$. Then, the allowed
initial values for the energy density are approximatively such
that~\cite{ZWP,BM1,BM2}: $10^{-37}\mbox{GeV}^4 \le \rho _Q \le 10^{61}
\mbox{GeV}^4$ where $10^{-37}\mbox{GeV}^4$ is the background energy
density at equality whereas $10^{61}\mbox{GeV}^4$ represents the
background energy density (\ie the radiation energy density) at the
initial redshift. If, for instance, $\alpha =6$ and if the scalar
field starts at rest, this means that the initial values of the field
are such that $10^{-18}\mP \le Q_{\rm in} \le 10^{-2}\mP$ just after
inflation. If $Q_{\rm in}$ is large initially, then the attractor is
joined quite recently. Of course, the range of allowed initial
conditions, when expressed in terms of the initial field, depends on
the value of the parameter $\alpha $.

\par 

The values of $\alpha $ are constrained by the measurement of the
equation of state today. The larger $\alpha $ is, the larger the
equation of state parameter $\omega _Q$ is today. Since it is known
that $\omega _Q$ cannot be too different from $-1$, this means that
$\alpha $ cannot be too large. In fact, this conclusion rests on the
particular shape of the Ratra-Peebles potential. From a model building
point of view, it is more natural to consider the SUGRA potential
given by~\cite{BM1,BM2,BMR1,BMR2}
\begin{equation}
W(Q)=W_0{\rm e}^{4\pi Q^2/\mP^2}\left(\frac{Q}{\mP}\right)^{-\alpha }\, .
\end{equation}
Then, the equation of state parameter today is modified. Since, today,
the value of the field is of the order of the Planck mass, the
supergravity exponential factor plays an important role in this
regime. This has two consequences. Firstly, the equation of state is
pushed toward the value $-1$ because the exponential factor increases
the importance of the potential energy with respect to the kinetic
energy. Secondly, the value of $\omega _Q$ becomes almost independent
of $\alpha $ because, again, the exponential factor dominates. As a
consequence the constraint on $\alpha $ can be relaxed and any value
is in fact a priori possible. Moreover, in the very early universe,
one has $Q\ll \mP$ and, this time, the exponential factor becomes
one. In this case, the SUGRA potential has the same shape as the
Ratra-Peebles potential. In summary, if we have the SUGRA model in
mind (which is, from a high-energy point of view, well-motivated),
then, in the early universe, we can safely work with the Ratra-Peebles
as an effective model (which is simpler, technically speaking) but
without the usual restrictions on the parameter $\alpha $.

\par

We have seen that the initial conditions are usually fixed just after
inflation, at the beginning of the radiation dominated era. In this
paper, we study the behavior of the quintessence field during the
phase of inflation itself. One of our main goal is to check whether
the behavior of $Q$ during inflation is compatible with the allowed
initial conditions (in fact, ``final'' conditions from the point of
view of the present study) described before. Clearly, if the final
value of $Q$ (at the end of inflation) is not in the allowed range,
then the quintessential scenario is in trouble. In fact, in order to
avoid the above-mentioned situation, our study rather helps us to put
constraints on the initial conditions, not at the beginning of the
radiation dominated era as before, but at the beginning of the
inflationary phase.

\par

We start with a study of the classical evolution of the quintessence
field (\ie without the quantum effects). The classical quintessence
field obeys the usual equation of motion for a scalar field in a FLRW
background with the Hubble parameter depending on the classical
inflaton field $\varphi_\cl$, namely
\begin{equation}
  {\ddot Q}_\cl+3H(\varphi_\cl)\dot Q_\cl + W'(Q_\cl)=0\, .
\end{equation}
Whenever the friction term is large, one can neglect the double
derivative term, obtaining the standard slow-roll equation
\begin{equation}
\label{QclassEq}
  \frac{{\rm d}Q_\cl}{{\rm d}t}+\frac{W'(Q_\cl)}{3H(\varphi_\cl)}=0\, .
\end{equation}
The consistency of this assumption can be directly checked. Taking the
time derivative of the last equation one can actually show that 
\begin{equation}
  \frac{\ddot Q_\cl}{H(\varphi_\cl)\dot Q_\cl}=
  -\frac{W''(Q_\cl)}{3H^2(\varphi_\cl)}+\epsilon,
\end{equation}
where $\epsilon=-\dot H/H^2$ is the usual inflaton slow-roll parameter
which is by assumption much smaller than $1$. Therefore, the slow roll
approximation can be applied to the equations describing the motion of
the quintessence field whenever the following condition is satisfied
\begin{equation}
  \frac{W''(Q_\cl)}{3H^2(\varphi_\cl)} \ll 1 \, .
\end{equation}
If one applies this condition to the Ratra-Peebles potential, one 
arrives at
\begin{equation}
\label{slowQ}
  \left(\frac{Q_{\rm cl}}{\mP}\right)^{\alpha+2} \gg
  \frac{\alpha(\alpha+1)W_0}{8\pi V(\varphi_\cl)}\, .
\end{equation}

\par

\begin{figure*}[t]
\begin{center}
  \includegraphics[width=.90\textwidth,height=.50\textwidth]{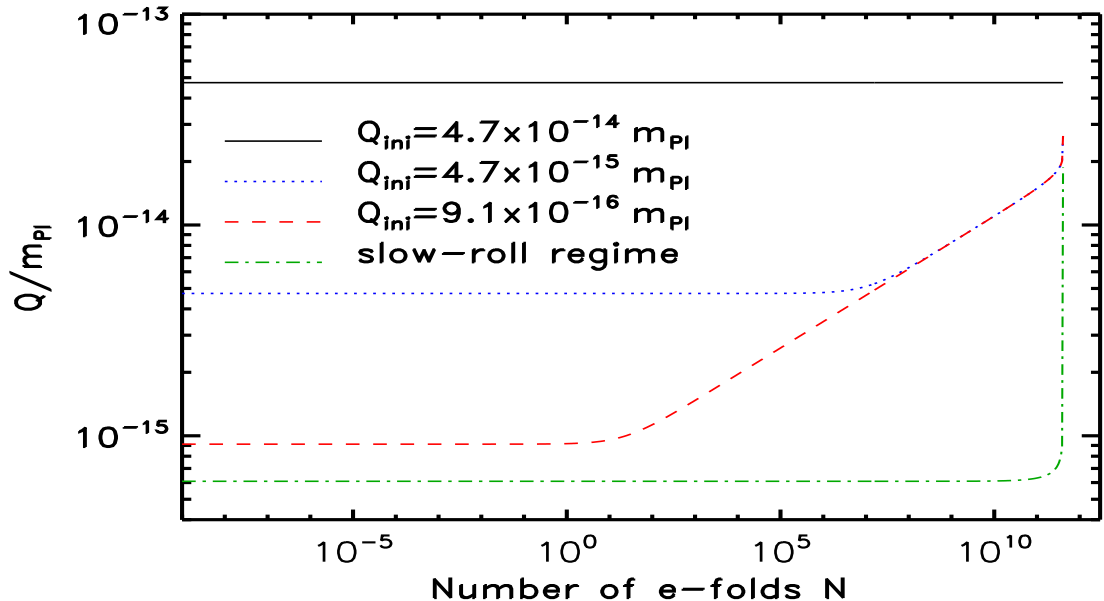} 
\caption{Classical evolution of the quintessence field for different
initial conditions. A massive chaotic inflaton potential (\ie $n=2$)
has been assumed and the parameter $\alpha $ in the Ratra-Peebles
potential has been taken to be equal to $6$. The initial condition for
the inflaton field is such that $\varphi _\ini=2.5\times 10^{6}\mP$
corresponding to an initial energy density of $V_\ini=0.1\mP^4$. The
green dotted-dashed line signals the limit of the slow-roll
approximation.  Below this line, the slow-roll approximation is no
longer valid.}
\label{Q}
\end{center}
\end{figure*}

Let us now turn to the solution of Eq.~(\ref{QclassEq}). For the
Ratra-Peebles potential given by Eq.~\eqref{ratra}, the solution to
this slow-roll equation can be expressed as
\begin{equation}
\label{QclassSol}
  \frac{Q_{\rm cl}}{\mP} = \left[\left(\frac{Q_\ini }{\mP}\right)^{\alpha +2}
  +\frac{\alpha(\alpha+2)W_0}{\mP ^2} 
  \int^{\varphi_\ini}_{\varphi_\cl(t)}\!\!  \frac{\de{0}{\chi
  }}{V'(\chi)}\right]^{1/(\alpha +2)}\, .
\end{equation}
Using the chaotic inflationary potential~\eqref{pot} we get for any
value of the power index such that $n>2$
\begin{multline}
\label{qsolnneq2}
  \frac{Q_\cl}{\mP} = \Biggl\{\left(\frac{Q_\ini }{\mP}\right)^{\alpha +2}
  + \frac{\alpha (\alpha +2)}{n(2-n)} \frac{W_0}{V_0}\\
  \times \left[\left(
\frac{\varphi_\ini}{\mP}\right)^{2-n}
-\left(\frac{\varphi_\cl}{\mP}\right)^{2-n}
\right]\Biggr\}^{1/(\alpha +2)}\, ,
\end{multline}
while, for $n=2$, the result reads
\begin{equation}
\label{qsoln2}
  \frac{Q_\cl}{\mP} = \left[\left(\frac{Q_\ini }{\mP}\right)^{\alpha +2}
  +\frac{\alpha (\alpha +2)}{2}\frac{W_0}{V_0}
\ln\frac{\varphi_\ini}{\varphi _{\rm cl}(t)}\right]^{1/(\alpha +2)}.
\end{equation}
These results are consistent with those obtained in
Ref.~\cite{ML}. The evolution of the quintessence field in the
slow-roll regime for different initial values is plotted in
Fig.~\ref{Q}.
Let us quickly discuss these solutions. One can obtain a more compact
expression if one notices that $\varphi _{\rm cl}\simeq \varphi _{\rm
in}[1-N/(2N_{_{\rm T}})]$ where $N_{_{\rm T}}$ is the total number of
e-folds during inflation. This expression is valid provided that
$N<4\pi (\varphi _{\rm in}/\mP)^2/n=N_{_{\rm T}}+n/4\simeq N_{_{\rm
T}}$. Then, one has
\begin{multline}
  \frac{Q_\cl}{\mP} \simeq \frac{Q_\ini}{\mP}
  \bigg[1+
  \frac{\alpha (\alpha +2)W_0}{2nV_0}
\left(\frac{\varphi _\ini}{\mP}\right)^{2-n}
\\
\times \left(\frac{Q_\ini}{\mP}\right)^{-\alpha -2}
\frac{N}{N_{_{\rm T}}}\bigg]^{1/(\alpha +2)}\, .
\end{multline}
If the initial value is large enough, then the dynamical term will
remain negligible and the field is frozen until inflation ends.  As a
matter of fact, for any value of $n$ we obtain that $Q_\cl(t)\simeq
Q_\ini$ at all times if the initial value satisfies the constraint
\begin{equation}
\label{constQini}
 \left( \frac{Q_\ini}{\mP}\right)^{\alpha +2}
\gtrsim\left(\frac{W_0}{V_0}\right)
\left(\frac{\varphi _\ini}{\mP}\right)^{2-n}\, .
\end{equation}
If the above condition is satisfied at initial time, then, obviously,
the condition given by Eq.~(\ref{slowQ}) is also satisfied. There is
also an intermediate regime for which the condition~(\ref{constQini})
is not satisfied but~(\ref{slowQ}) is. In this case, the field is not
frozen, even at initial times. Finally, if $Q_\ini$ is sufficiently
small so that the condition~(\ref{slowQ}) is violated, then we expect
the quintessence dynamics to rapidly bring back the field to the
slow-roll regime, where the previous considerations apply.

\subsection{Perturbative Solutions}

As mentioned at the beginning of this article, in order to study the
evolution of the quintessence field, it is not sufficient to integrate
the classical equation of motion since the quantum effects can play an
important role and modify the classical evolution. We now analyze the
stochastic behavior of the quintessence field, in the case where the
total energy density is still dominated by the vacuum energy of the
inflaton field. The stochastic evolution of the quintessence field $Q$
is controlled by a Langevin equation which, in the slow-roll
approximation, reads
\begin{equation}
\label{Langevin_quint}
  \frac{{\rm d}Q}{{\rm d}t}+\frac{W'(Q)}{3H(\varphi)} =
  \frac{H^{3/2}(\varphi)}{2\pi}\xi_Q(t)\, ,
\end{equation}
where $\xi_Q$ is another white-noise field such that
\begin{equation}
  \mean{\xi_Q(t)\xi_Q(t')}=\delta(t-t'),\quad
  \mean{\xi_Q(t)\xi(t')}=0\, .
\end{equation}
The solution of the Langevin equation~(\ref{Langevin_quint}) depends
explicitly on $\xi_Q$ but also on the inflaton noise $\xi$ through the
coarse-grained field $\varphi$.

\par

In order to find an approximate solution to Eq.~\eqref{Langevin_quint}, one
may try to use the same perturbative technique as the one used before
for the inflaton field. Therefore, we expand the quintessence field
about the classical slow-roll solution~\eqref{QclassSol} and write
\begin{equation}
  Q(t)=Q_{\rm cl}(t)+\delta Q_1+\delta Q_2 + \cdots \, .
\end{equation}
Then, it is easy to establish that the equations of motion for the
perturbed quantities $\delta Q_1$ and $\delta Q_2$ are given by the
following expressions
\begin{widetext}
\begin{eqnarray}
\frac{{\rm d}\delta Q_1}{{\rm d}t}+\frac{W''(Q_{\rm cl})}{3H(\varphi
_{\rm cl})}\delta Q_1 &=& \frac{W'(Q_{\rm cl})H'(\varphi _{\rm
cl})}{3H^2(\varphi _{\rm cl})}\delta \varphi _1
+\frac{H^{3/2}(\varphi _{\rm cl})}{2\pi }\xi _Q\, , 
\\
\frac{{\rm d}\delta Q_2}{{\rm d}t}+\frac{W''(Q_{\rm cl})}{3H(\varphi
_{\rm cl})}\delta Q_2 &=& \frac{W'(Q_{\rm cl})H'(\varphi _{\rm
cl})}{3H^2(\varphi _{\rm cl})}\delta \varphi _2
+\frac{W'(Q_{\rm cl})H''(\varphi _{\rm
cl})}{6H^2(\varphi _{\rm cl})}\delta \varphi _1^2
-\frac{W'(Q_{\rm cl})H'^2(\varphi _{\rm
cl})}{3H^3(\varphi _{\rm cl})}\delta \varphi _1^2
\nonumber \\
& & +\frac{W''(Q_{\rm cl})H'(\varphi _{\rm
cl})}{3H^2(\varphi _{\rm cl})}\delta \varphi _1\delta Q_1
-\frac{W'''(Q_{\rm cl})}{6H(\varphi _{\rm cl})}\delta Q_1^2
+\frac{3H^{1/2}(\varphi _{\rm cl})H'(\varphi _{\rm
cl})}{4\pi }\delta \varphi _1\xi _Q\, .
\end{eqnarray}
\end{widetext}
Although these equations look quite complicated, they can be solved
easily because (by definition) they are linear. The solution for
$\delta Q_1$ reads
\begin{multline}
\delta Q_1(t)=W'(Q_{\rm cl})\int _\tin ^t \biggl[\frac{H'(\varphi _{\rm
cl})}{3H^2(\varphi _{\rm cl})}\delta \varphi_1 (\tau ) \\
+\frac{H^{3/2}(\varphi _{\rm cl})}{2\pi W'(Q_{\rm cl})}\xi _Q(\tau
)\biggr] {\rm d}\tau \, ,
\end{multline}
and, as required, is linear both in the quintessence noise $\xi_Q$ and
(through $\delta\varphi_1$) in the inflaton noise $\xi$. As a
consequence, $\delta Q_1$ has a vanishing mean value
\begin{equation}
  \mean{\delta Q_1}=0 \, , 
\end{equation}
but a non-vanishing variance given by the sum of two contributions
originating from the inflaton and quintessence noise variances, namely
\begin{widetext}
\begin{eqnarray}
\label{varq11}
\mean{\delta Q_1^2} &=&\frac{W'^2(Q_{\rm cl})}{9}
\int _\tin^t \int _\tin^t \frac{H'(\tau )}{H^2(\tau )}
\frac{H'(\eta )}{H^2(\eta )}
\mean{\delta \varphi _1(\tau )\delta \varphi _1(\eta )}
{\rm d}\tau {\rm d}\eta 
\nonumber \\
& & 
+\frac{W'^2(Q_{\rm cl})}{4\pi ^2}
\int _\tin^t \int _\tin^t \frac{H^{3/2}(\tau )}{W'(\tau )}
\frac{H^{3/2}(\eta )}{W'(\eta )}
\mean{\xi _Q(\tau )\xi _Q(\eta )}
{\rm d}\tau {\rm d}\eta 
\\
\label{varq12}
&\equiv & 
\mean{\delta Q_1^2}\vert _{\xi _{\varphi}}+
\mean{\delta Q_1^2}\vert _{\xi _{Q}}\, .
\end{eqnarray}
Let us notice that there is no mixed contribution since the
cross-correlation $\mean{\xi \xi_Q}=0$. The detailed calculation of
$\mean{\delta Q_1^2}$, in particular its explicit expression in terms
of the inflaton field and/or the number of e-folds $N$, is rather
lengthy and is carried out in the Appendix~\ref{pertQ}.

\par

Let us now turn to the second order correction. The solution for
$\delta Q_2$ can be written as
\begin{eqnarray}
\label{sol2}
\delta Q_2(t) &=& W'(Q_{\rm cl})\int _\tin ^t \biggl\{\frac{H'(\varphi _{\rm
cl})}{3H^2(\varphi _{\rm cl})}\delta \varphi_2 (\tau )
+\left[\frac{H''(\varphi _{\rm
cl})}{6H^2(\varphi _{\rm cl})}
-\frac{H'^2(\varphi _{\rm
cl})}{3H^3(\varphi _{\rm cl})}\right]\delta \varphi_1^2(\tau )
+\frac{W''(Q_{\rm cl})H'(\varphi _{\rm
cl})}{3W'(Q_{\rm cl})H^2(\varphi _{\rm cl})}
\delta \varphi_1(\tau )\delta Q_1(\tau )
\nonumber \\
& & -\frac{W'''(Q_{\rm cl})}{6W'(Q_{\rm cl})H(\varphi _{\rm cl})}
\delta Q_1^2(\tau )
+\frac{3}{4\pi }
\frac{H^{1/2}(\varphi _{\rm
cl})H'(\varphi _{\rm
cl})}{W'(Q_{\rm cl})}\delta \varphi _1(\tau )\xi _{Q}\biggr\}
{\rm d}\tau \, .
\end{eqnarray}
As expected, one sees that $\delta Q_2$ is quadratic in the
noises. From the above expression, one deduces that the mean value of
$\delta Q_2$ is non-vanishing and is the sum of various terms
\begin{equation}
\label{meanq2}
\mean{\delta Q_2}=
\mean{\delta Q_2}\vert _{\delta \varphi _2} 
+\mean{\delta Q_2}\vert _{\delta \varphi _1^2} 
+\mean{\delta Q_2}\vert _{\delta \varphi _1\delta Q_1} 
+\mean{\delta Q_2}\vert _{\delta Q_1^2(\xi _Q)}
+\mean{\delta Q_2}\vert _{\delta Q_1^2(\xi _\varphi)}\, ,
\end{equation} 
\end{widetext}
where the last term in Eq.~(\ref{sol2}) does not contribute because 
\begin{equation}
\mean{\delta \varphi _1\xi _{Q}}=0\, .
\end{equation}
If we had not taken into account the stochastic behavior of the
inflaton, only the term $\mean{\delta Q_2}\vert _{\delta Q_1^2(\xi
_Q)}$ would have contributed. Again, the explicit expressions of each
term are given in the Appendix~\ref{pertQ}.

\par

Let us now quickly present what the outcome of the perturbative
approach applied to the evolution of the quintessence field is. The
main result is that when the classical evolution of the quintessence
field is negligible, \ie when the condition \eqref{constQini} holds,
the variance reads
\begin{equation}
\label{MLvar}
\mean{\delta Q_1^2}\vert _{\xi _{Q}}=
\frac{16\mP^2}{3n(n+2)}\frac{V_0}{\mP^4}\left[\left(\frac{\varphi
_\ini}{\mP}\right)^{n+2}-\left(\frac{\varphi
_\cl}{\mP}\right)^{n+2}\right]\, .
\end{equation}
which is the same result already obtained in Ref.~\cite{ML} that we
recover here with a different method.

\begin{figure*}[t]
  \includegraphics[width=.90\textwidth,height=.50\textwidth]{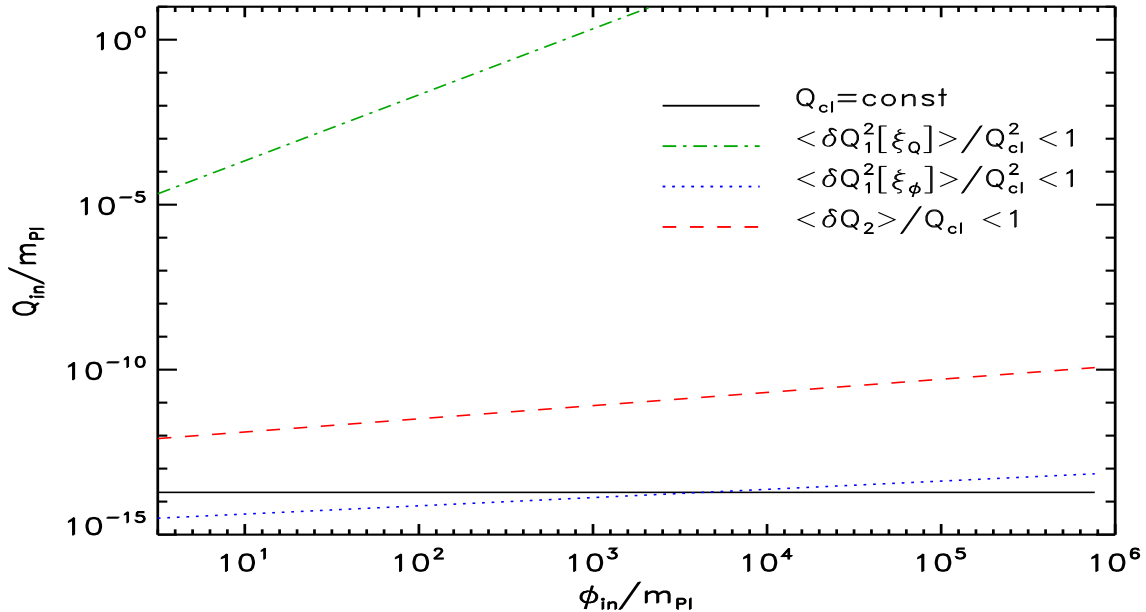}
\caption{\label{constraint} Constraints coming from the requirement
that the perturbative approach is valid for the case $n=2$ and $\alpha
=6$. The values of the initial conditions $\varphi _\ini $ and $Q_\ini
$ are chosen such that the corresponding energy density is never
Planckian, \ie $V(\varphi _\ini )<\mP^4 $ and $V(Q_\ini)<\mP^4$. Each
line corresponds to a condition specified in the figure and that we
describe below. By convention, the allowed region is always the region
above the corresponding line. The solid black line indicates the
regime where $Q_\cl $ can be considered as constant, see
Eq.~(\ref{constQini}). The dot-dashed green line corresponds to the
region where $\mean{\delta Q_1^2}/Q_\cl ^2<1$, originating from the
quintessence noise, is small, see Eq.~(\ref{condvarqless1}). The
dotted blue line corresponds to the same condition except that the
variance now comes from the inflaton noise. Finally, the dashed red
line corresponds to the condition $\mean{\delta Q_2}/Q_\cl <1$ where
the mean value originates from the quintessence noise. As we have seen
before, see Eq.~(\ref{meanq2}), there are four other contributions but
one can easily show that they do not lead to new constraints and this
is the reason why they are not represented in this figure. As
discussed in the text, the most stringent constraint comes from the
requirement $\mean{\delta Q_1^2}/Q_\cl ^2<1$.}
\end{figure*}

Another general result that is established in details in the
Appendix~\ref{pertQ} is that, when the perturbative approach is valid,
the contribution coming from the inflaton noise is completely
negligible, see for instance Figs.~\ref{perturbative}.
 
\par

Unfortunately, this is not a very strong argument because it is
possible to see that the perturbative treatment is well under control
only in a very small region of the parameter space. Indeed, one must
check that the conditions
\begin{equation}
  \frac{\mean{\delta Q_1^2}}{Q_\cl^2} \ll 1 \, , \quad 
  \frac{\mean{\delta Q^2}}{Q_\cl} \ll 1\, .
\end{equation}
are fulfilled if one wants the perturbative approach to be valid. This
leads to several constraints that are summarized in
Fig.~\ref{constraint}. More precisely, from Eq.~(\ref{varq12}), one
sees that the first condition leads to two constraints while, looking
at Eq.~(\ref{meanq2}), one notices that the second one leads to fives
constraints. As is apparent from Fig.~\ref{constraint}, the most
stringent constraint comes from the fact that the variance originating
from the quintessence noise must be small in comparison with
$Q_\cl^2$. Explicitly, working out this condition, it boils down to
\begin{equation}
\label{condvarqless1}
  \frac{Q_\ini}{\mP} \gtrsim
  \frac{\sqrt{V_0}}{\mP^2}\left(\frac{\varphi_\ini}{\mP}\right)^{n/2+1}\,
  .
\end{equation}
Let us also notice that this condition is a much more stringent
condition than \eqref{constQini}, see Fig.~\ref{constraint}. For most
values of $Q_\ini$ and $\varphi_\ini$, we rapidly get that
$\mean{\delta Q_1^2}/Q_\cl^2 \gg 1 $, meaning that a large part of the
initial condition space cannot be described by means of the
perturbative approach.  Therefore, there is the need for a different
approach.

\subsection{The Reflecting Wall}

The failure of the perturbative treatment is a signal of the fact that
the classical evolution is not a good zeroth order solution for the
perturbative expansion. The reason is that the equation of motion is
dominated by the diffusive term due to the noise while the classical
drift is in fact sub-dominant. It is therefore natural to neglect the
classical term and to solve the approximate equation
\begin{equation}
  \frac{{\rm d}Q}{{\rm d}t} = \frac{H^{3/2}(\varphi)}{2\pi}\xi\, .
\end{equation}
The solution to this equation can be written as
\begin{equation}
\label{solwall}
  Q=Q_\ini + \int_\tin^t\de{.5}{\tau}\frac{H^{3/2}(\varphi)}{2\pi}\xi(\tau)\, .
\end{equation}
As a first step, we also neglect the inflaton fluctuations and take
$\varphi$ to be the classical field $\varphi_\cl$. As a consequence,
$Q$ has a Gaussian probability distribution the mean of which is given
by $\mean{Q}=Q_\ini$ with a variance $\sigma _Q^2\equiv
\mean{Q^2}-\mean{Q}^2$ which can be expressed as
\begin{eqnarray}
\label{initvar}
& &\sigma _Q^2=\sigma_0^2(t)\equiv \int_\tin^t\de{.5}{\tau}
\frac{H^3(\varphi_\cl)}{4\pi^2}\nonumber \\ &=&
\frac{16\mP^2}{3n(n+2)}\frac{V_0}{\mP^4}\left[\left(\frac{\varphi
_\ini}{\mP}\right)^{n+2}-\left(\frac{\varphi
_\cl}{\mP}\right)^{n+2}\right]\, .
\end{eqnarray}
One recognizes Eq.~(\ref{MLvar}), first obtained in Ref.~\cite{ML},
and that we have re-derived in the preceding subsection by means of the
perturbative approach. Here, we have shown that Eq.~(\ref{MLvar}) can
be valid even when the perturbative approach breaks down.

\par

At this point a remark on the notation is in order. In the following
$\sigma _Q^2$ always denotes the variance of the quintessence field
while $\sigma _0^2$ is just a function defined by the above expression
which turns out to be equal to $\sigma _Q^2$ in the situation where
the classical drift and the inflaton noise are neglected.

\par

The above model is in fact too simple for the following reason. The
function $\sigma_0(t)$ increases with time and, at some point, is
larger than $Q_\ini$. In this case, there is a finite probability
(tending to 50\% at late times) that the quintessence field becomes
negative. Clearly, this is not possible because, in this case, the
classical term in the Langevin equation becomes dominant and prevents
$Q$ to become negative. In other words, in this regime, the classical
drift cannot be neglected. So, it seems that we are in fact back to
the original problem which consists in solving exactly the full
Langevin equation. However, there is a simple way out. Indeed, we can
model the effect of the classical term by considering that there is a
perfectly reflecting wall at $Q=0$ the purpose of which is of course
to prevent the quintessence field to become negative. As discussed in
Ref.~\cite{chandra}, the probability distribution of a random walk is
modified by a reflecting barrier in a way that is easy to
estimate. Let us assume that we start with a normalized probability
distribution for the variable $x$, $\int _{-\infty }^{+\infty}P(x){\rm
d}x=1$. Then, let us put a reflecting wall at $x=a$ such that only the
values $x>a$ are allowed. Then, Ref.~\cite{chandra} tells us that the
new probability distribution is $P(x)+P(2a-x)$ and it is easy to check
by mean of a simple change of variable that it is indeed normalized,
\ie $\int _{a}^{+\infty} [P(x)+P(2a-x)]{\rm d}x=1$. In the present
context, we have $a=0$ and, hence, the resulting probability
distribution becomes, for $Q>0$,
\begin{equation}
\label{pdfQ}
  P(Q,t)=\frac{\mathrm{e}^{-(Q-Q_\ini)^2/(2\sigma_0^2)} + 
  \mathrm{e}^{-(Q+Q_\ini)^2/(2\sigma_0^2)}}{\sqrt{2\pi}\sigma_0(t)}\, ,
\end{equation}
where $\sigma _0$ is given by Eq.~(\ref{initvar}) which means that,
for the moment, the contribution coming from the inflaton noise is
still neglected.

\par

As mentioned before, the advantage of the reflecting wall model is
that it permits simple analytical estimates of the relevant physical
quantities. However, there is one feature of the model that is worth
stressing here. The classical drift term which prevents the field to
become negative depends on the parameter $\alpha $ but the wall, which
is supposed to model this term, don't. Therefore, one limitation of
the reflecting wall model is that we have lost the $\alpha
$-dependence of the result. Concretely, in the following, we will see
that the mean value and/or the variance of $Q$ are $\alpha $
independent. Only an exact solution (or a numerical calculation) could
allow us to test the accuracy of this assumption. 

\par

With the help of this probability distribution we can now calculate
the mean and the variance of the quintessence field. For the mean, one
obtains the following analytical expression
\begin{equation}
\label{meanwall}
  \mean{Q} = \sigma_0
  \left[\sqrt{\frac{2}{\pi}}\mathrm{e}^{-Q_\ini^2/2\sigma_0^2}
  + \frac{Q_\ini}{\sigma_0}
  \mathrm{Erf}\left(\frac{Q_\ini}{\sqrt{2}\sigma_0}\right)\right]\, ,
\end{equation}
where $\mathrm{Erf}(z)$ is the error function defined by
$\mathrm{Erf}(z)\equiv (2/\sqrt{\pi }) \int _0^z{\rm d}t {\rm
e}^{-t^2}$. To our knowledge, this explicit formula is new and has not
been given elsewhere. In the same manner, one has $\mean{Q^2} =
\sigma_0^2 + Q_\ini^2$ and, therefore, one has
\begin{equation}
\label{varwallwonoise}
\sigma _Q^2=\sigma_0^2 + Q_\ini^2-\mean{Q}^2\, ,
\end{equation}
where $\mean{Q}$ is given by Eq.~(\ref{meanwall}). One notices the
variance of the quintessence field $\sigma _Q^2$ is now different from
the function $\sigma _0^2$, $\sigma _Q^2\neq \sigma _0^2$. Let us
emphasize again that the above results are not subject to the
limitations of the perturbative approach. In particular, the
difference between the classical value and the quantum (or stochastic)
average needs not to be small.

\begin{figure*}[t]
\includegraphics[width=.90\textwidth,height=.50\textwidth]{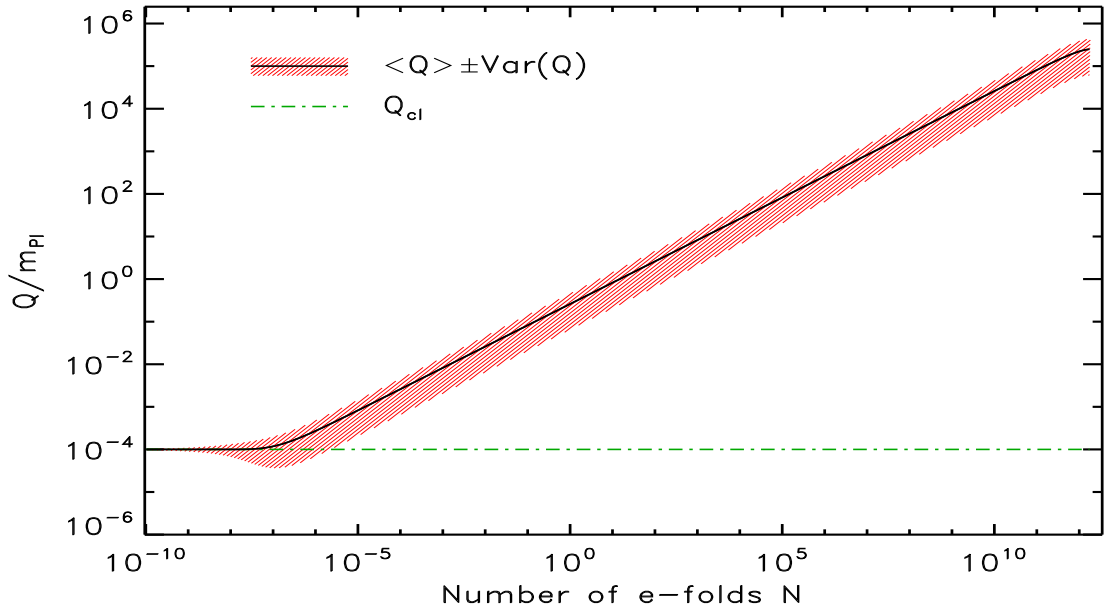}
\caption{Evolution of $\mean{Q}$ (solid black line) calculated from
Eq.~(\ref{meanwall}) (but with the inflaton noise taken into account)
and compared with the corresponding classical evolution (dotted-dashed
green line) $Q_\cl$. The dashed area represents the zone $\pm \sigma $
around the mean value where $\sigma $ has been obtained with the help
of Eq.~(\ref{varwall}). The difference between $\mean{Q}$ and $Q_\cl$
at the end of inflation can be many orders of magnitude (and is, in
the present case, of about $8$ orders of magnitude) demonstrating that
it is crucial to take into account the quantum effects during
inflation. The curves have been obtained for a massive chaotic model,
\ie $n=2$ and for the Ratra-Peebles potential with $\alpha =6$. The
initial value of the inflaton is $\varphi _\ini=5.4\times 10^{6}\mP$
corresponding to an initial energy density of $V_\ini=0.5\times
\mP^4$. The initial value of the quintessence field is
$Q_\ini=10^{-4}\mP$ corresponding to $W_\ini=1.25\times
10^{-99}\mP^4$.}
\label{cc}
\end{figure*}

\begin{figure*}[t]
\includegraphics[width=.90\textwidth,height=.50\textwidth]{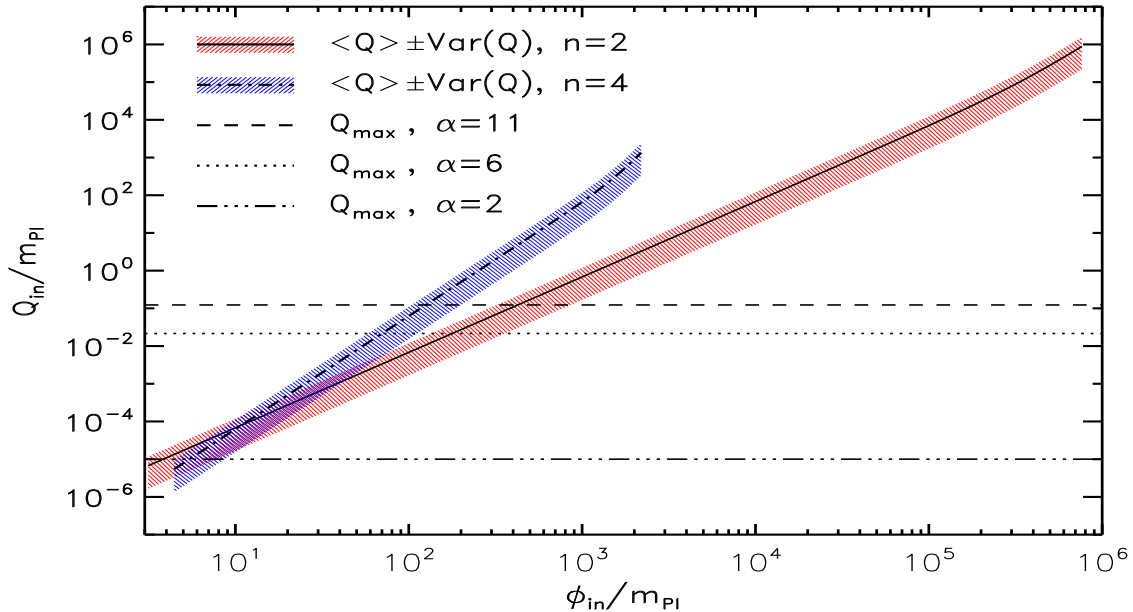}
\caption{\label{ciquint} The quantity $\mean{Q}$ at the end of
inflation versus the initial value of the inflaton field calculated
according to Eq.~(\ref{finalQ}) for two chaotic models, $n=2$ and
$n=4$. As noticed in the text, $\mean{Q}$ at the end of inflation does
not depend on the initial value of the quintessence field because
there is an attractor solution given by Eq.~(\ref{attra}). The fact
that the final value does not depend on the parameter $\alpha $ is due
to the description of the classical drift term by means of the
reflecting wall. The two curves start and stop at different values of
the initial inflaton field because the minimal value (defined to be
the one which leads to at least $60$ e-folds) and the maximal value
(defined to be the one which corresponds to $V_\ini=\mP^4$) are not
the same according to the model (\ie $n$) under considerations (for
instance, $\varphi_\ini=3.1\mP$ for $n=2$ while $\varphi _\ini=4.4\mP$
for $n=4$). The horizontal lines represent the maximal values of
$\mean{Q}$ (for different values of the index $\alpha $) at the end of
inflation such that the attractor is joined today. These values are
computed by means of Eq.~(\ref{safety}).}
\end{figure*}

\par

If we insert Eq.~(\ref{initvar}) into Eqs.~(\ref{meanwall}) and
(\ref{varwallwonoise}), one obtains the mean value $\mean{Q}$ and the
variance $\sigma _Q^2$ as functions of the inflaton and/or the number
of e-folds. At initial time, one has $\sigma _0\rightarrow 0$ and
therefore $\mean{Q}\rightarrow Q_\ini$, where we have used the fact
that $\mathrm{Erf}(z)\rightarrow 1$ when $z\rightarrow +\infty$. For
the variance, one has in the same regime (\ie initially for $\varphi
\rightarrow \varphi _\ini$ and $\sigma_0(t)\ll Q_\ini$)
\begin{equation}
  \sigma _Q^2\rightarrow \sigma_0^2(t)\, , 
\end{equation}
where we have used Eq.~(\ref{varwallwonoise}) and the behavior of the
mean at initial times. This means that, in this regime, we recover the
previous results obtained by means of the perturbative approach.

\par

One can also study what happens at late times when $\sigma_0(t)\gg
Q_\ini$. In this case, one has
\begin{equation}
\label{attra}
  \mean{Q} \rightarrow \sqrt{\frac{2}{\pi }}\sigma_0(t)\, .
\end{equation}
The situation is reminiscent to the behavior of quintessence in the
more traditional situation where the background evolution is dominated
either by radiation or matter. Indeed, as it is the case in this
well-studied context and as explained before, the late time evolution
of $\mean{Q}$ is independent on the initial conditions, \ie on
$Q_\ini$. In other words, there is an attractor for $\mean{Q}$ given
by Eq.~(\ref{attra}). In this situation, the final value of the
quintessence field only depends on the initial value of the inflaton
field (and, of course, on which kind of potential is responsible for
inflation: in the present context, it only depends on $n$). However,
this conclusion should be toned down because we will see in the
following that another quantity of interest, namely the probability
that the quintessence field be on tracks today (or, equivalently, that
its value at the end of inflation be in a given range) does depend on
$Q_\ini$ (and, in fact, also on $\varphi _\ini$).

\par

Using the above expression, the variance of the quintessence field at
late times can also be estimated. It is given by
\begin{equation}
  \sigma _Q^2\rightarrow \left(1-\frac{2}{\pi
  }\right)\sigma_0^2(t)\, .
\end{equation}
We see that, as $\sigma _0(t)$ increases and as the probability that
the random-walking field is reflected by the wall becomes non
negligible, the mean value $\mean{Q}$ is shifted toward higher values
while its variance slightly shrinks.

\par

We are now in a position where we can come back to one of our starting
questions, namely the influence of the inflaton noise. In particular,
we study whether including the inflaton fluctuations can cause
significant deviations from the above results and, if so, under which
physical conditions this is the case. In order to take into account
this effect, we re-start from Eq.~(\ref{solwall}). In this formula,
the argument of the Hubble parameter is no longer the classical
inflaton but the stochastic process studied in the previous section.
Then, one can use the perturbative treatment studied before but, and
this is of course the crucial point, only for the inflaton
field. Indeed, we have seen before that this perturbative treatment
(contrary to the perturbative treatment for the quintessence field) is
almost always reliable. Performing a Taylor expansion of
Eq.~(\ref{solwall}), one obtains
\begin{widetext}
\begin{eqnarray}
\label{solwallinf}
Q &=& Q_\ini + \frac{1}{2\pi }\int_\tin^t\de{.5}{\tau}
\biggl\{
H^{3/2}(\varphi _{\rm cl})+\frac32H'(\varphi _{\rm cl})
H^{1/2}(\varphi _{\rm cl})\delta \varphi _1(\tau )
+\frac32H'(\varphi _{\rm cl})
H^{1/2}(\varphi _{\rm cl})\delta \varphi _2(\tau )
\nonumber \\
& & +\frac12\biggl[\frac32H''(\varphi _{\rm cl})
H^{1/2}(\varphi _{\rm cl})
+\frac34H'^2(\varphi _{\rm cl})
H^{-1/2}(\varphi _{\rm cl})\biggr]\delta \varphi _1^2(\tau )
+\cdots  \biggr\}\xi(\tau)\, .
\end{eqnarray}
\end{widetext}
Equipped with this solution, one can now reintroduce the reflecting
wall and recalculate the various moments. Clearly, since the solution
is still linear in the quintessence noise, the probability function of
the quintessence field, taking into account the wall, is still given
by Eq.~(\ref{pdfQ}) but, in the expression of $P(Q,t)$, $\sigma _0$
should now be replaced by another function (of the classical inflaton
noise or of the number of e-folds) that, in the following, we simply
denote by $\sigma $ (not to be confused with $\sigma _Q$). The
replacement of $\sigma _0$ by $\sigma $ is the only change
needed. Otherwise the expression of the new $P(Q,t)$ is similar to the
one given by Eq.~(\ref{pdfQ}). Let us now determine $\sigma $
explicitly. As before, $\sigma $ is simply given by the variance
deduced from Eq.~(\ref{solwallinf}) using the fact that we have white
noises. Eq.~(\ref{solwallinf}) implies that $\sigma ^2$ is given by
\vspace{-2mm}
\begin{equation}
  \sigma ^2(t) = 
  \int_\tin^t\de{.5}{\tau}\frac{\mean{H^3(\varphi)}}{4\pi^2}\, ,
\end{equation}
where the mean value in the integral can be expressed through a power
expansion of $H^3(\varphi)$ up to second order about its classical
value yielding
\begin{multline}
  \mean{H^3(\varphi)}=H^3(\varphi_\cl)
  +  3H^2H'\mean{\dphi_2}\\
  +  3\left[H(H')^2 +\frac{H^2H''}{2}\right]\mean{\dphi_1^2}\, ,
\end{multline}
that can now be easily integrated. The final result is
\nopagebreak{
\begin{widetext}
\vspace{-5mm}
\begin{eqnarray}
\label{varwall}
  \sigma^2(t) &=& \sigma_0^2(t) + 
  \frac{16\mP^2}{3n}\frac{V_0^2}{\mP^8}\Biggl\{ \frac{4}{3n}
\left(\frac{\varphi_\ini}{\mP}\right)^{n/2+2}
\left[\left(\frac{\varphi_\ini}{\mP}\right)^{3n/2} 
- \left(\frac{\varphi _\cl}{\mP}\right)^{3n/2}\right]+
\left(\frac{\varphi_\ini}{\mP}\right)^4\left[
\left(\frac{\varphi_\ini}{\mP}\right)^{2n-2} 
- \left(\frac{\varphi _\cl}{\mP}\right)^{2n-2}\right] 
\nonumber \\
& & -\frac{n}{n+1}\left[\left(\frac{\varphi_\ini}{\mP}\right)^{2n+2} - 
\left(\frac{\varphi _\cl}{\mP}\right)^{2n+2}\right]
  \Biggr\}\, 
\end{eqnarray}
\vspace{-1mm}
\end{widetext}
\onecolumngrid
\vspace*{-4mm} \twocolumngrid 
Let us now compare the function $\sigma (N)$ with the function $\sigma
_0 (N)$. At early times, \ie when $\varphi_\cl(t)\sim \varphi_\ini$, a
linearization is sufficient in order to obtain a good approximation
(and this is in fact the case for a large part of inflation). A
striking feature of the above result is that the extra term coming
from the inflaton quantum fluctuations cancel out, the first
non-vanishing contribution being of order $\mathcal{O}[(N/N_{_{\rm
T}})^2]$, where we remind that $N_{_{\rm T}}$ is the total number of
e-folds. Therefore, for the main part of the inflationary era we can
safely consider that $\sigma(t)\simeq\sigma_0(t)$. Again, we reach the
conclusion that the effects originating from the inflaton noise do not
play a crucial role. Conversely, at late times, when
$\varphi_\cl(t)\ll\varphi_\ini$, the above expression yields
\begin{equation}
\label{latesigma}
  \sigma^2(t)=\sigma_0^2(t)\left[1+\frac{(n+2)(7n+4)}{3n(n+1)}
\frac{V_\ini}{\mP^4}\right]\, ,
\end{equation}
meaning that, as could be expected, the contribution of the inflaton
quantum fluctuations is significant only when inflation starts near
the Planck scale, while it becomes negligible for smaller values of
the initial energy density.

\par

We are now in a position where one can compute the new mean and
variance exactly. As already mentioned, the new probability function
is given by Eq.~(\ref{pdfQ}) with $\sigma _0$ replaced by $\sigma
$. This immediately means that $\mean{Q}$ and $\sigma _Q^2$ are given
by Eqs.~(\ref{meanwall}) and (\ref{varwallwonoise}) with $\sigma _0$
replaced by $\sigma $. The evolution of $\mean{Q}$ versus the number
of e-folds is displayed in Fig.~\ref{cc}. At initial times, one has
$\mean{Q}\simeq Q_\ini$ and $\sigma _Q^2\simeq \sigma ^2\simeq \sigma
_0^2$, the last approximate equality coming from the property
established before Eq.~(\ref{latesigma}). On the other hand, the mean
value and the variance of the quintessence field at late times read
(if the initial condition for the inflaton field is large enough in
order to reach the regime where $\sigma\gg Q_\ini$)
\begin{equation}
\label{finalQ}
\mean{Q}= \sqrt{\frac{2}{\pi}}\;\sigma \, ,\quad 
\sigma _Q= \sqrt{1-\frac{2}{\pi}}\;\sigma
\end{equation}
where in the above equations $\sigma $ is the $\sigma(t)$ calculated at the
end of inflation, which is given by
\begin{equation}
  \sigma = \varphi_\ini \sqrt{\frac{16}{3n(n+2)}
  \frac{V_{\rm in}}{\mP^4} \left[1+\frac{(n+2)(7n+4)}{3n(n+1)}
  \frac{V_\ini}{\mP^4}\right]}\, ,
\end{equation}
where we have used Eq.~(\ref{latesigma}) and an approximation for
$\sigma _0$ valid at late times. One notices that the two quantities
of Eq.~(\ref{finalQ}) growing function of $\varphi_\ini$. This can be
checked explicitly in Fig.~\ref{ciquint} where the final values of
$\mean{Q}$ versus $\varphi_\ini$ are represented.

\par

There is another important consequence that can be deduced from what
has been discussed so far. As already mentioned, the energy density of
the quintessence field at the beginning of the radiation era (\ie at
the end of inflation) must be such that $10^{-113}\mP^4<\rho
_Q<10^{-9}\mP^4$ if one wants $Q$ to be on tracks today. Assuming for
simplicity that the field starts at rest, this implies
\begin{equation}
\label{safety}
\frac{Q_{\rm min}}{\mP}\equiv 10^{-107/\alpha } <\frac{Q}{\mP}
<10^{-9/\alpha }\equiv \frac{Q_{\rm max}}{\mP}\, .
\end{equation}
Since, during inflation, $Q$ is a stochastic quantity, one can study
the probability for a given model (that is to say for a given choice
of the power indices and of the initial conditions $Q_\ini$ and
$\varphi_\ini$) that the field is in the appropriate range. This
probability allows us to evaluate the likelihood of the various models
under considerations, rejecting those for which this quantity is
small. In this way, we can thus constrain the value of the initial
conditions for the fields and exclude a portion of the parameter
space. Two remarks are in order here. Firstly, as is clear from
Eq.~(\ref{pdfQ}), the probability will depend on $Q_\ini$. Therefore,
although the late time evolution of $\mean{Q}$ is independent of
$Q_\ini$, this dependence is reintroduced via the calculation of the
probabilities. Secondly, the result will also depend on $\varphi
_\ini$. As a consequence, for a fixed value of $n$ and $\alpha $, one
can hope to derive constraints on the initial value of the
quintessence field but also on the total number of e-folds during
inflation.

\par

A given model will be accepted if a large part of its probability
distribution calculated at the end of inflation is contained within
the allowed range. A rough estimate of this constraint can be obtained
by simply requiring that the mean value of the distribution falls in
this range within one square root of the variance.  Using the previous
results of this section, one finds that imposing that
$\mean{Q}\pm\sigma _Q$ falls between $Q_{\rm min}$ and $Q_{\rm max}$
yields
\begin{equation}
  Q_{\rm min}<\left(\sqrt{\frac{2}{\pi}}\pm\sqrt{1-\frac{2}{\pi}}\right)\sigma
  < Q_{\rm max}\, .
\end{equation}
Therefore, in order to obtain a probability of order one must have
that $\sigma\lesssim Q_{\rm max}$. This implies an upper bound on
$\varphi_\ini$ that can be roughly estimated to (neglecting the
contribution from inflaton fluctuations)
\begin{equation}
  \frac{\varphi_\ini}{\mP} \lesssim
  \left(\frac{\mP^2}{\sqrt{V_0}}\frac{Q_{\rm
  max}}{\mP}\right)^{2/(n+2)} \sim 10^{10(\alpha-2)/[\alpha(n+2)]}\, .
\end{equation}
It is easy to see from the above formula that this constraint is quite
stringent. Let us also notice that this also a constraint on the total
number of e-folds during inflation, very roughly speaking $N_{_{\rm
T}}\lesssim 10^{20(\alpha-2)/[\alpha(n+2)]}$.

\par

More precisely, if one uses the probability density function given by
Eq.~(\ref{pdfQ}) (with, as already discussed at length, $\sigma _0$
replaced by $\sigma $ if one wants to take into account the inflaton
noise at the end of inflation), one can calculate the exact
probability for the quintessence to be on tracks today. One arrives at
\begin{widetext}
\begin{eqnarray}
\label{probatracks}
P(Q_{\rm min}<Q<Q_{\rm max}\vert Q_\ini,\alpha, \phi _\ini, n)
&=&
\frac12 \biggl[{\rm Erf}\left(\frac{Q_{\rm max}-Q_\ini}{\sqrt{2}\sigma}\right)
-{\rm Erf}\left(\frac{Q_{\rm min}-Q_\ini}{\sqrt{2}\sigma}\right)
+{\rm Erf}\left(\frac{Q_{\rm max}+Q_\ini}{\sqrt{2}\sigma}\right)
\nonumber \\
& & -{\rm Erf}\left(\frac{Q_{\rm min}+Q_\ini}{\sqrt{2}\sigma}
\right)\biggr]\, .
\end{eqnarray}
\end{widetext}

\begin{figure*}
\psfrag{!}{\!\!/}
\includegraphics[width=.47\textwidth,height=.40\textwidth]{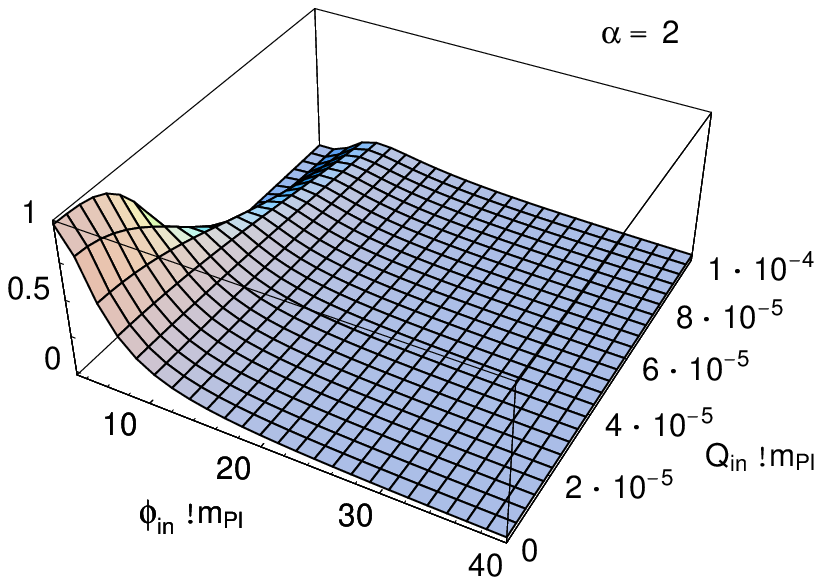}
\includegraphics[width=.47\textwidth,height=.40\textwidth]{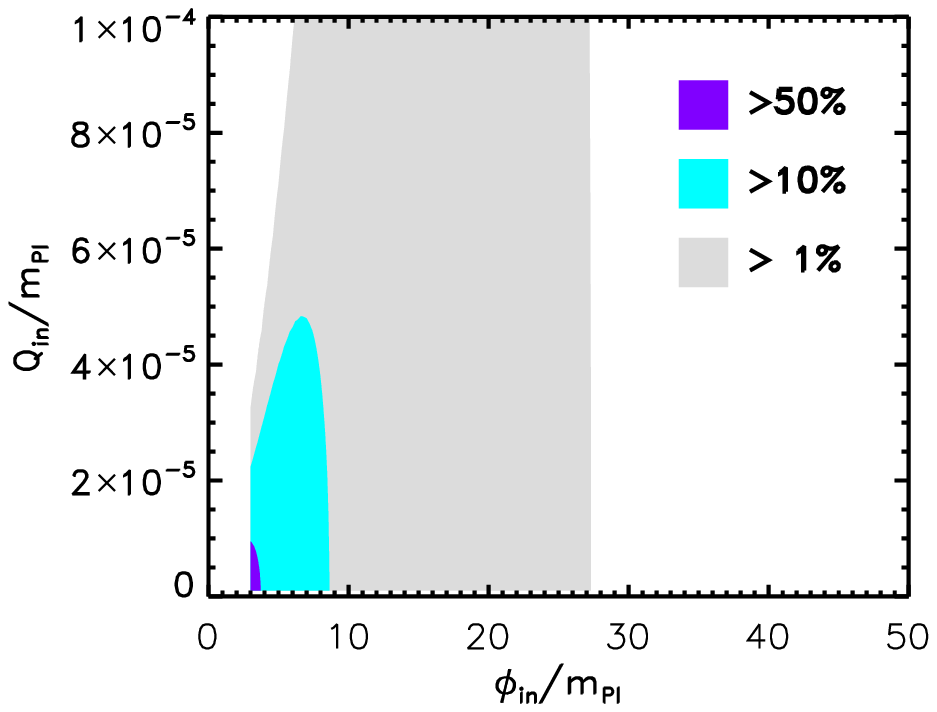}
\\
\includegraphics[width=.47\textwidth,height=.40\textwidth]{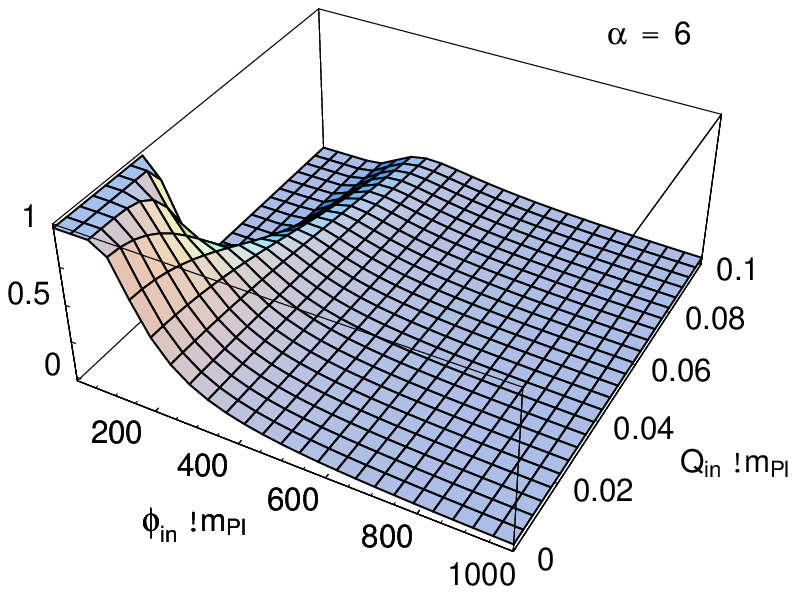}
\includegraphics[width=.47\textwidth,height=.40\textwidth]{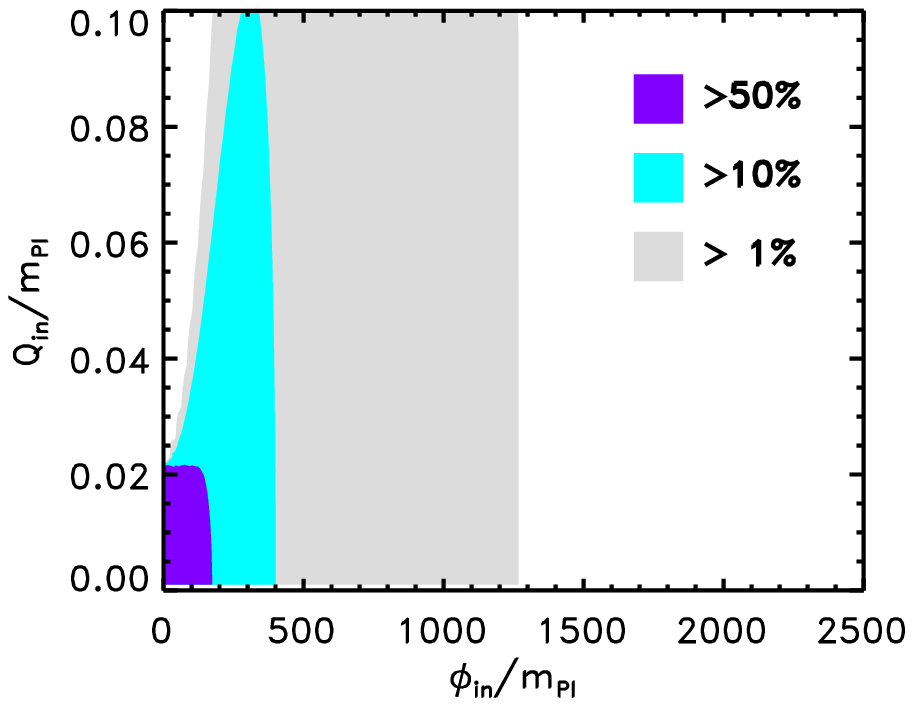}
\\
\includegraphics[width=.47\textwidth,height=.40\textwidth]{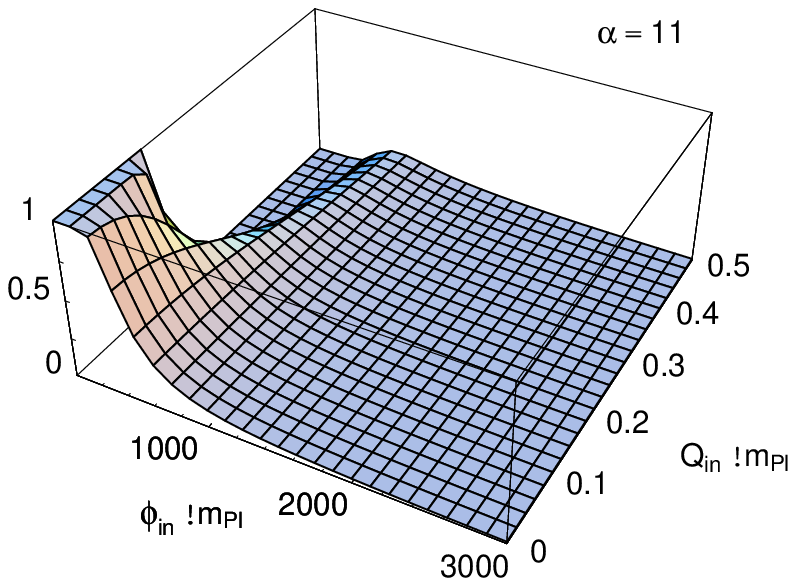}
\includegraphics[width=.47\textwidth,height=.40\textwidth]{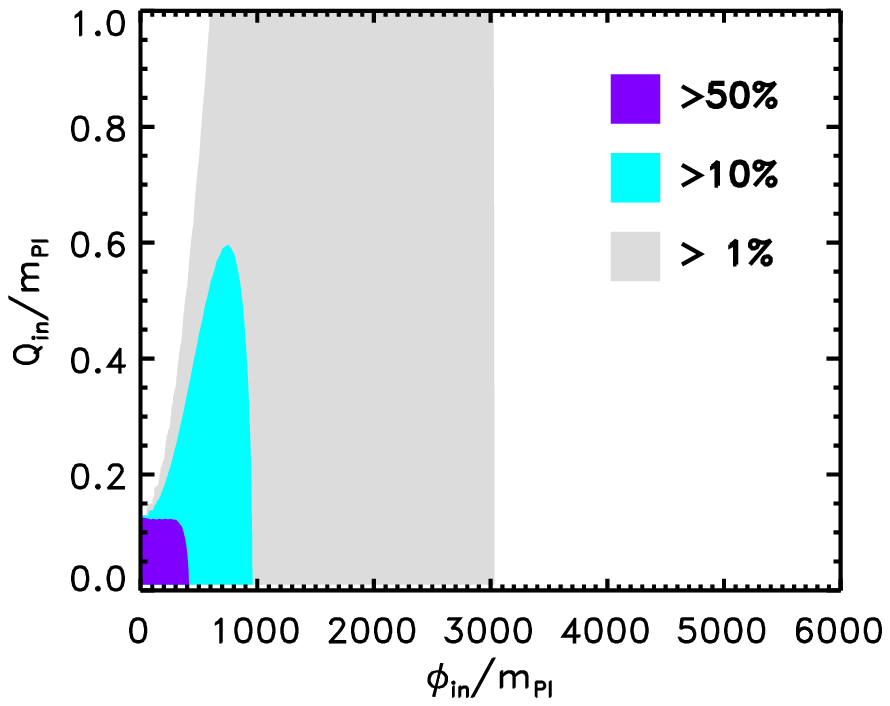}
\caption{\label{proba_init_cond} Probability computed from
Eq.~(\ref{probatracks}) that the quintessence field is on tracks today
for various models with $n=2$ and different power indices $\alpha$
(from top to bottom, $\alpha =2,6,11$). For all the models under
consideration, the maximum allowed value of the initial inflaton is
$\varphi_\ini=7.6\times 10^5 \mP$ (coming from the fact that
$V_\ini<\mP^4$). From the plots, one notices that values larger than
$27\mP$, $1.3\times 10^3 \mP$ and $3.0\times 10^3 \mP$ respectively
(corresponding to an initial energy density $V_\ini$ of $1.3\times
10^{-9} \mP^4$, $2.9\times 10^{-6} \mP^4$ and $1.5\times 10^{-5}
\mP^4$) are excluded at $99\%$ CL. One also notices that, since the
confidence region enlarges for increasing values of $\alpha$, large
values of the power index $\alpha $ are statistically preferred.}
\end{figure*}

The detailed behavior of this probability as a function of the
initial conditions $\varphi_\ini$ and $Q_\ini$ is shown in
Fig.~\ref{proba_init_cond} for $n=2$ and different power indices
$\alpha$.  We see that a large portion of the parameter space can be
excluded. Actually, for $n=2$, the largest allowed initial condition
is $\varphi_\ini\simeq 7.6\times 10^5 \mP$, corresponding to the
Planck energy density. However, at $99\%$ of confidence level, initial
values larger than $27\mP$ ($\alpha=2$), $1.3\times 10^3 \mP$
($\alpha=6$) or $3.0\times 10^3 \mP$ ($\alpha=11$) are excluded.  This
also shows that, since the confidence region enlarges with the power
index $\alpha$, large values of $\alpha$ are statistically more
favored than small values.

\section{Discussion and Conclusions}

In this section, we briefly discuss other aspects of the question
studied in this article and present our conclusions.

\par

In the preceding sections, we have calculated the evolution of the
quintessence field during inflaton taking into account the quantum
effects. What about these quantum effects in the subsequent
cosmological eras? After the reheating, during the radiation and
matter dominated phases, it is easy to show that the quintessence
field evolves classically starting from the value reached at the end
of inflation and determined by its random walk during
inflation. Indeed, since the noise in the Langevin equation is
controlled by the quantum fluctuations of the modes leaving the
horizon, it rapidly becomes negligible as the accelerated expansion
stops and the modes start to reenter the horizon.

\par

However, a new phase of accelerated expansion driven by the
quintessence field is now taking place and, clearly, the above
argument does not apply in this case.  Therefore, one may wonder
whether the influence of the quantum effects should not be taken into
account when one computes the evolution of $Q$ at present time. This
could have important observational consequences, in particular if the
stochastic behavior of $Q$ modifies the value of the equation of
state. However, it is easy to demonstrate that this is not so. The
problem is in fact very similar to calculating the evolution of the
inflaton field during inflation since, at present time, the
quintessence field is no longer a test field but actually determines
the evolution of the background. Therefore, even though the slow roll
approximation might not be so satisfactory in this case, we can at
least estimate the relevance of these late stochastic fluctuations by
simply setting $n=-\alpha$ into~Eq.~\eqref{varinflaton} (since, in
some sense, the Ratra-Peebles potential is nothing but an inflationary
chaotic potential with a negative index). This equation, provided one
substitutes $\varphi _\cl$ with $Q_\cl$ and $V(\varphi _\cl)$ with
$W(Q_\cl)$, should be a reasonable estimate of $\mean{\delta
Q^2}/Q^2$. Since the quintessence field is now on the attractor, its
value must be of the order of the Planck mass. In addition, it is
slowly rolling down its potential toward higher values. As a
consequence, from Eq.~\eqref{varinflaton}, one gets the following
rough estimate
\begin{equation}
  \frac{\mean{\delta Q^2}}{Q^2}\bigg\vert_\mathrm{now} \lesssim
  \frac{W(Q_{\rm now})}{\mP^4}\simeq 10^{-123}\, ,
\end{equation}
meaning that any stochastic deviation from the classical trajectory is
completely negligible at present time.

\par

Let us now end this work by reviewing what are the main
conclusions of our study. The main result is that, taking into account
the quantum effects during inflation is important since the stochastic
diffusion term dominates over the classical drift term and that, as a
consequence, the value of $\mean{Q}$ can differ from $Q_\cl$ by
several orders of magnitude. For the first time to our knowledge, we
have given an analytical estimate describing the evolution of
$\mean{Q}$ during inflation, see Eq.~(\ref{meanwall}).

\par

Another new result is the fact that, requiring the quintessence field
to have a large probability to be on tracks today, allows us to put
quite stringent constraints on the initial conditions $Q_\ini$ and
$\varphi _\ini$. Typically, the quintessence field must start from
small values. We have also established that large values of $\alpha $
are favored (we notice in passing that, if we consider the
Ratra-Peebles potential only and not the SUGRA one, the same
conclusion is reached from the constraints that exist on the equation
of state today). Another interesting result is that the inflaton field
must also start from quite small values. This implies that the total
number of e-folds during inflation is also limited. On the other hand,
we have remarked the existence of an attractor for $\mean{Q}$, see
Eq.~\eqref{attra}, due to the fact that the final value of $\mean{Q}$
is independent of $Q_\ini$. However, a dependence in the initial
conditions is reintroduced in the calculation of the probability which
has allowed us to put the constraints mentioned just before.

\par

One of the main purpose of our paper was also to study the influence
of the inflaton noise on the evolution of the quintessence field. The
approximation consisting in neglecting the inflaton fluctuations has
been shown to be justified in most cases, basically because the
corresponding contributions to the mean value and/or to the variance
are proportional to $V_\ini/\mP^4$, see for instance
Eq.~(\ref{finalQ}). Even in the extreme case of Planckian initial
conditions for the inflaton field (\ie $V_\ini\sim\mP^4$), the
inflaton noise is unlikely to modify $\mean{Q}/\mP$ by more than one
order of magnitude compared to what is obtained taking into account
the quintessence noise only.

\par

It is also interesting to compare these results to those obtained in
the paper~\cite{ML} which was the first to take into account the
quantum effects in the calculation of the evolution of
$\mean{Q}$. Basically, our findings confirm and/or justify the results
of Ref.~\cite{ML} and somewhat extend their validity. We have
recovered the same equation for the variance and our new
equation~(\ref{meanwall}) for the mean value of $Q$ confirms the
conclusions that can be drawn from the figures of Ref.~\cite{ML},
namely that the quantum effects can play an important role during
inflation. In Ref.~\cite{ML}, the inflaton noise has not been
considered and, as mentioned above, we have demonstrated that this is,
in most cases, a good approximation.

\par

Finally, let us describe some questions that are left unanswered and
some possible improvements to the present study. In order to be able
to find analytical solutions, we have modeled the classical drift
term with a reflecting wall. The price to pay is that we have lost the
dependence in the parameter $\alpha $. The drift term acts differently
for different Ratra-Peebles potentials while the wall repels the field
regardless to $\alpha $. Although we do not expect a strong
dependence, it would be interesting to quantify this effect. The
problem is that, if one includes the exact classical term, then the
Langevin equation is no longer analytically solvable. The only way out
seems to numerically integrate this equation. However, even this
solution could be difficult because a term like $Q^{-\alpha }$ can
rapidly become very large and, hence, problematic from the numerical
point of view.

\par

Another interesting question would be to study what happens when one
considers the case of a colored noise since it is clear that a white
noise is not, physically, the most relevant case. Concretely, this
amounts to replace the Heaviside function in the expansion of the
field by a smooth function and, in principle, this could affect the
evolution of the quintessence field during inflation although, again,
we do not expect a very important effect.

\par

For the moment, we postpone the study of all the issues to future
works.

\vspace{0.5cm} 
\centerline{\bf Acknowledgments}
\vspace{0.2cm}

We wish to thank S.~Matarrese for many enlightening comments and
discussions.

\appendix

\section{Reflecting Wall for the Inflaton}
\label{wallinf}

Another way to look at the quantum evolution of the inflaton is the
following. From Eq.~(\ref{varnewnoise}), it is clear that the variance
of $\Psi (t)$ is a growing function of time. However, in order for the
solution of Eq.~(\ref{solphi}) to be defined we need to impose the
condition $\Psi(t)<1$. Otherwise, this equation is clearly meaningless
and, of course, if this condition is not satisfied, the series of
Eq.~(\ref{solserieslambda}) is not convergent. This is probably the
reason for the problems encountered before. The above condition can be
thought of as constraining the random walk of $\Psi$ with a reflecting
wall. As explained in the preceding section, see also
Ref.~\cite{chandra}, the resulting probability distribution for $\Psi$
with the wall located at $\Psi =1$ is given by
\begin{equation}
\label{pdfpsi}
P(\Psi,t)= \frac{{\rm e}^{-\Psi^2/\left(2\mean{\Psi^2}\right)} +
{\rm e}^{-(\Psi-2)^2/\left(2\mean{\Psi^2}\right)}}{\sqrt{2\pi
\mean{\Psi^2(t)}}}\, ,
\end{equation}
and the corresponding probability distribution for $\varphi$, obtained
via the relation $P[\varphi,t]=P[\Psi(\varphi),t]\vert {\rm
d}\Psi/{\rm d}\varphi \vert$, becomes
\begin{equation}
P(\varphi,t)=P\left[\Psi(\varphi)\right]\frac{2}{\varphi
_\cl}\left(\frac{\varphi _\cl}{\varphi }\right)^3\, .
\end{equation}
This probability distribution has a finite mean value that can be
expressed as [this equation can also be obtained by using the link
between $\Psi $ and $\varphi $ given by Eq.~(\ref{solphi}) and the
probability distribution of $\Psi$ given by Eq.~(\ref{pdfpsi})]
\begin{equation}
  \mean{\varphi}=\varphi_\cl \frac{{\rm
  e}^{-1/\left(4\mean{\Psi^2}\right)}}{\sqrt{\mean{\Psi^2}}}
  \sqrt{\frac{\pi}{2}}
  I_{-1/4}\!\!\left(\frac{1}{4\mean{\Psi^2}}\right),
\end{equation}
where $I_\nu(x)$ is the modified Bessel function of the first
kind. For small values of $\mean{\Psi^2}$, the mean value reads
\begin{equation}
  \mean{\varphi}=\varphi_\cl\left[1+\frac{3}{8}\mean{\Psi^2}
  +\mathcal{O}\left(\mean{\Psi^2}^2\right)\right],
\end{equation}
yielding therefore the same results as~\eqref{solserieslambda}. Let
also notice that all higher moments are divergent in accordance with
the discussion presented in the section on the evolution of the
inflaton field.
\vspace{5mm} 

\section{Perturbative solution for the quintessence field}
\label{pertQ}

\subsection{Solution at First Order}

In this appendix, we present the explicit expressions (as a function
of the classical inflaton field and/or of the number of e-folds) of
$\mean{\delta Q_1^2}$ given by Eqs.~(\ref{varq11}) and~(\ref{varq12})
and of $\mean{\delta Q_2}$ given by Eq.~(\ref{meanq2}). We start with
Eq.~(\ref{varq12}) which is the sum of two terms. In order to evaluate
the first term coming from the inflaton noise, \ie $\mean{\delta
Q_1^2}\vert _{\xi _{\varphi}}$, one must calculate the two-point
correlation function of $\delta \varphi _1$. Using the solution of
Eq.~(\ref{soldphi1}), one obtains
\begin{widetext}
\vspace{-5mm}
\begin{equation}
\mean{\delta \varphi _1(\tau )\delta \varphi _1(\eta )} =\frac{1}{4\pi
^2}H'(\tau)H'(\eta )\left[\int _\tin ^{\eta }\frac{H^3(\sigma
)}{H'^2(\sigma )} {\rm d}\sigma +\Theta (\eta -\tau )
\int _\eta ^{\tau }\frac{H^3(\sigma
)}{H'^2(\sigma )} {\rm d}\sigma \right]\, .
\end{equation}
In the above expression, $\Theta (z)$ is the Heaviside function, \ie
is zero if $z<0$ and one otherwise. Then, using the link between the
cosmic time and the classical inflaton field in the slow-roll
approximation, the remaining integrations can be easily performed
since they just boil down to integrating power-law functions. One
obtains
\begin{eqnarray}
\frac{\mean{\delta Q_1^2}}{\mP^2}\biggl\vert _{\xi _{\varphi}} &=&
\frac{4\alpha ^2}{3n^2}\frac{W_0}{V_0}
\frac{W_0}{\mP^4}\left(\frac{Q_\cl}{\mP}\right)^{-2\alpha -2}
\Biggl\{\frac{1}{8-n} \left[\left(\frac{\varphi _\cl
}{\mP}\right)^{-n/2}-\left(\frac{\varphi _{\rm
in}}{\mP}\right)^{-n/2}\right]\times \left[\left(\frac{\varphi _\cl
}{\mP}\right)^{-n/2+4}-\left(\frac{\varphi _{\rm
in}}{\mP}\right)^{-n/2+4}\right] \nonumber 
\\ 
& & +\frac{1}{n} \left(\frac{\varphi _{\rm in}}{\mP}\right)^4
\left[\left(\frac{\varphi _\cl}{\mP}\right)^{-n/2}-\left(\frac{\varphi
_{\rm in}}{\mP}\right)^{-n/2}\right]\times \left[\left(\frac{\varphi
_\cl}{\mP}\right)^{-n/2}-\left(\frac{\varphi _{\rm
in}}{\mP}\right)^{-n/2}\right] \nonumber
\\ 
& & -\frac{4}{(4-n)(8-n)}
\left[\left(\frac{\varphi _\cl}{\mP}\right)^{4-n}-\left(\frac{\varphi
_{\rm in}}{\mP}\right)^{4-n}\right] -\frac{1}{8-n} \left(\frac{\varphi
_{\rm in}}{\mP}\right)^{-n/2+4} \left[\left(\frac{\varphi
_\cl}{\mP}\right)^{-n/2}-\left(\frac{\varphi _{\rm
in}}{\mP}\right)^{-n/2}\right] \nonumber \\ & & +\frac{1}{8-n}
\left(\frac{\varphi _{\rm in}}{\mP}\right)^{-n/2}
\left[\left(\frac{\varphi _\cl}{\mP}\right)^{-n/2+4}-\left(\frac{\varphi
_{\rm in}}{\mP}\right)^{-n/2+4}\right]\Biggr\}\, .
\end{eqnarray}
The evolution of $\mean{\delta Q_1^2}\vert _{\xi _{\varphi}}$ as a
function of the number of e-folds is displayed in
Fig.~\ref{perturbative}.  The main feature of this formula is that it
is proportional to $(W_0/V_0) \times (W_0/\mP^4)$ which is, as
discussed before, very small. This is confirmed by the plot, see the
left panel in Fig.~\ref{perturbative}.
\begin{figure*}[t]
\includegraphics[width=.49\textwidth,height=.55\textwidth]{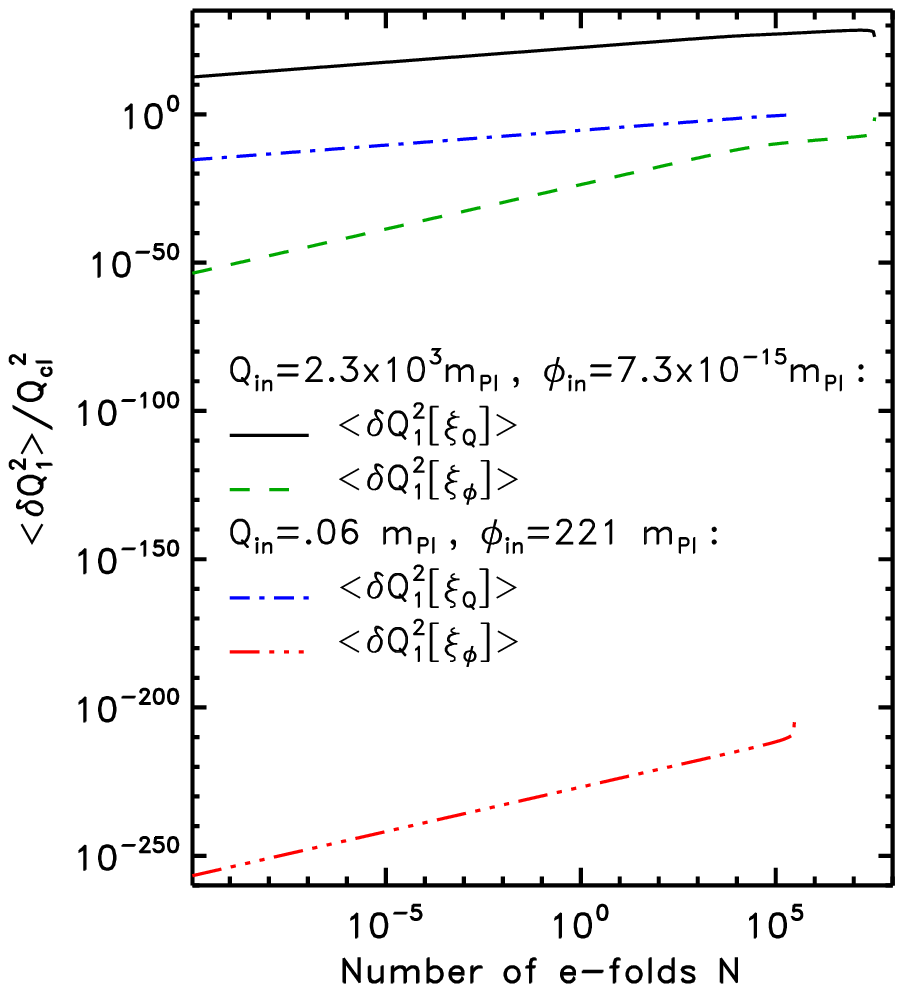}
\includegraphics[width=.49\textwidth,height=.55\textwidth]{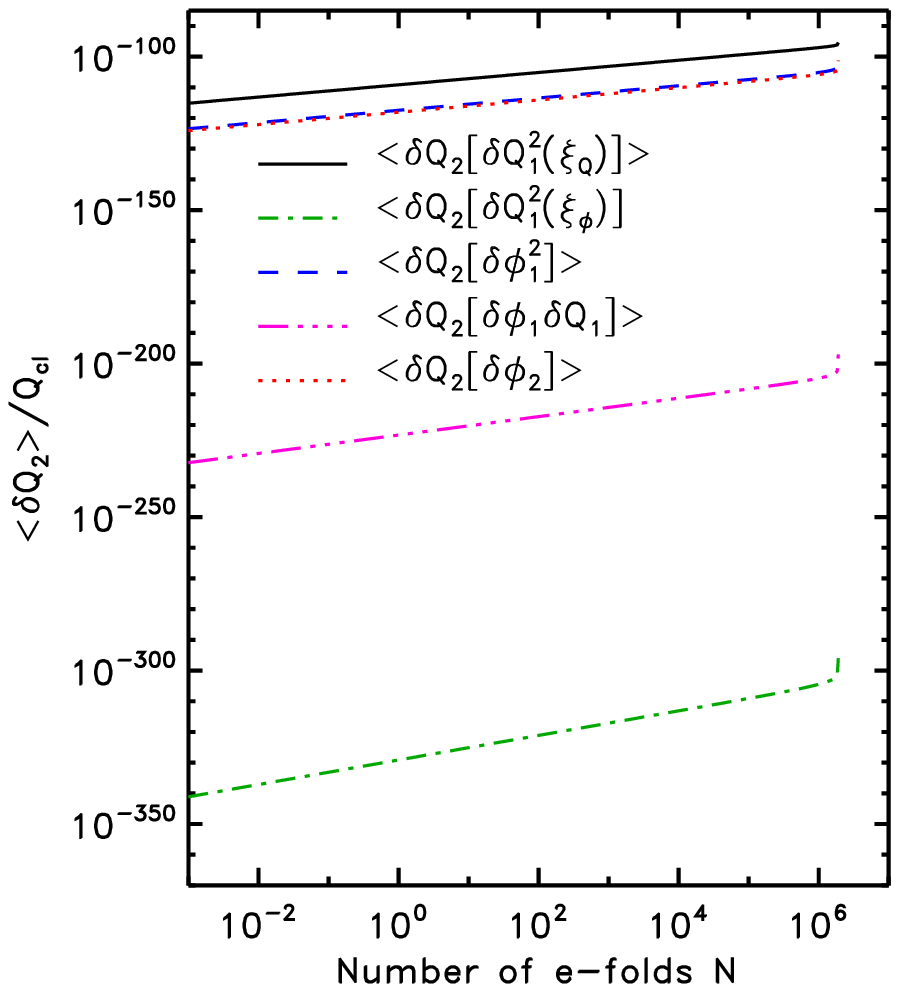}
\caption{Left panel: evolution of $\langle \delta
Q_1^2\rangle/Q_\cl^2$ calculated in the perturbative regime for two
different set of initial conditions (\ie two different values of
$\varphi _\ini$ and $Q_\ini$) and for a model with $n=2$ and $\alpha
=6$. For each set of initial conditions specified explicitly in the
figure, the contribution coming from the quintessence noise and the
contribution originating from the inflaton noise are represented. The
perturbative treatment is under control only in the case
$\varphi_\ini=221 \mP$ and $Q_\ini= 0.06 \mP$ corresponding to initial
energy densities given by $V_\ini=3.7\times 10^{-8} \mP^4$ and
$W_\ini=2.8\times 10^{-114} \mP^4$ (since all the corresponding curves
lie in the region $\langle \delta Q_1^2\rangle/Q_\cl^2<1$). The plot
demonstrates clearly that, in the perturbative regime, the inflaton
noise is totally negligible. Right panel: evolution of $\langle \delta
Q_2\rangle/Q_\cl$ for the initial conditions $\varphi_\ini=221 \mP$
and $Q_\ini= 0.06 \mP$. The five curves correspond to the five
contributions of Eq.~(\ref{meanq2}). Again, it is obvious that the
four contributions originating from the inflaton noise are
unimportant.}
\label{perturbative}
\end{figure*}

\par

Let us now turn to the term sourced only by the quintessence
noise. Using the fact that the quintessence noise is white the second
term of Eq.~(\ref{varq11}) becomes
\begin{equation}
\mean{\delta Q_1^2}\vert _{\xi _{Q}}=\frac{W'^2(Q_{\rm cl})}{4\pi ^2}
  \!\!\int_\tin^t\de{.5}{\tau} \frac{H^3(\varphi _{\rm
  cl})}{W'^2(Q_{\rm cl})},
\end{equation}
Then, using the expression of the inflaton and quintessence
potentials, straightforward calculations lead to the following formula
in terms of hypergeometric functions
\begin{eqnarray}
\label{q12xiq}
\frac{\mean{\delta Q_1^2}}{\mP^2}\biggl\vert _{\xi _{Q}} &=&
\frac{16}{3n(2-n)\mu}\frac{V_0}{\mP^4} \left(\frac{Q_\cl}{Q_{\rm
in}}\right)^{-2\alpha -2}\times \Delta ^{-\nu}
\nonumber \\
&\times &\Biggl\{ 
\left(\frac{\varphi _\cl}{\mP}\right)^{\mu (n-2)}
{}_2F_1\left[\nu,\mu,\mu +1,\frac{\Delta 
(\varphi _{\rm in}/\mP)^{2-n}-1}{\Delta }
\left(\frac{\varphi _\cl}{\mP}\right)^{n-2}\right]
\nonumber \\
&- &
\left(\frac{\varphi _{\rm in}}{\mP}\right)^{\mu (n-2)}
{}_2F_1\left[\nu,\mu,\mu +1,\frac{\Delta 
(\varphi _{\rm in}/\mP)^{2-n}-1}{\Delta }
\left(\frac{\varphi _{\rm in}}{\mP}\right)^{n-2}\right]
\Biggr\}\, ,
\end{eqnarray}
where we have used the definitions
\begin{equation}
\mu \equiv \frac{n+2}{n-2}-\frac{2(\alpha +1)}{\alpha +2}\, ,\quad 
\nu \equiv -\frac{2(\alpha +1)}{\alpha +2}\, ,\quad \Delta \equiv 
\frac{\alpha (\alpha +2)}{n(n-2)}\Upsilon\, ,\quad \Upsilon
\equiv \frac{W_0}{V_0}
\left(\frac{\mP}{Q_{\rm in}}\right)^{\alpha +2}\, ,
\end{equation}
and where we have assumed that $n\neq 2$. The number $\mu $ is always
positive for $n\le 6$ which is the case we are mostly interested in
and the number $\nu $ is, on the contrary, always negative. The
evolution of $\mean{\delta Q_1^2}\vert _{\xi _{Q}}$ is represented in
Fig.~\ref{perturbative}. The presence of the hypergeometric functions
is linked to the fact that the classical quintessence field is not
totally frozen. Looking at Eq.~(\ref{qsolnneq2}), one sees that the
term responsible for the slight evolution of $Q_\cl$ is in fact
proportional to $\Delta $. Indeed, in the limit $\Delta \rightarrow
0$, or equivalently $\Upsilon \rightarrow 0$, the classical
quintessence field becomes $Q=Q_{\rm in}$, \ie is actually frozen. In
this case, one has
\begin{equation}
{}_2F_1\left[\nu,\mu,\mu +1,\frac{\Delta (\varphi _{\rm
in}/\mP)^{2-n}-1}{\Delta } \left(\frac{\varphi
_\cl}{\mP}\right)^{n-2}\right]\rightarrow 
\left(\frac{\varphi _\cl}{\mP}\right)^{-\nu (n-2)}\Delta ^{\nu }
\frac{\mu }{\mu -\nu}\, ,
\end{equation}
and Eq.~(\ref{q12xiq}) reads
\begin{equation}
\frac{\mean{\delta Q_1^2}}{\mP^2}\biggl\vert _{\xi _{Q}} \rightarrow 
\frac{16}{3n(n+2)}\frac{V_0}{\mP^4}
\left[\left(\frac{\varphi _{\rm ini}}{\mP}\right)^{n+2}
-\left(\frac{\varphi _\cl}{\mP}\right)^{n+2}\right]\, ,
\end{equation}
which is exactly the expression derived in Ref.~\cite{ML}. This term
is proportional to the factor $V_0/\mP^4$ which is much larger than
$(W_0/V_0) \times (W_0/\mP^4)$. As a result, we expect the term
originating from the quintessence noise to dominate over the term
originating from the inflaton noise and, as already mentioned, this
conclusion is confirmed by the plots in Fig.~\ref{perturbative}. A
priori this conclusion is valid only in the regime where the above
expressions have been established, \ie in the perturbative regime.

\par

We have noticed before that the solution expressed in terms of the
hypergeometric function is valid provided $n\neq 2$. The case $n=2$,
for which the evolution of the classical quintessence field is given
by Eq.~(\ref{qsoln2}), requires a special treatment. In this case, one
finds
\begin{eqnarray}
\label{q12xiqn=2}
\frac{\mean{\delta Q_1^2}}{\mP^2}\biggl\vert _{\xi _{Q}} &=&
\frac{16}{3\times 8}\frac{V_0}{\mP^4} \left(\frac{Q_\cl}{Q_{\rm
in}}\right)^{-2\alpha -2}
\left(\frac{\varphi _{\rm in}}{\mP}\right)^{4}
\left[\frac{8}{\alpha (\alpha +2)\Upsilon }\right]^{\nu }
\exp\left[\frac{8}{\alpha (\alpha +2)\Upsilon }\right]
\nonumber \\
&\times & \left\{ 
\gamma \left[\frac{3\alpha +4}{\alpha +2}, \frac{8}{\alpha (\alpha
+2)\Upsilon }-4\ln \frac{\varphi _\cl}{\varphi _{\rm in}}\right]
-\gamma \left[\frac{3\alpha +4}{\alpha +2}, \frac{8}{\alpha (\alpha
+2)\Upsilon }\right]\right\} \, ,
\end{eqnarray}
where $\gamma (\beta ,x)$ is the incomplete gamma function defined by
\begin{equation}
\gamma (\beta ,x)\equiv \int _0^x{\rm e}^{-t}t^{\beta -1}{\rm d}t\, .
\end{equation}
In the limit $\Upsilon \rightarrow 0$, using the formula $\gamma
(\beta ,x)\simeq \Gamma (\beta )-x^{\beta -1}{\rm e}^{-x}$ when
$x\rightarrow 0$, one obtains
\begin{eqnarray}
\frac{\mean{\delta Q_1^2}}{\mP^2}\biggl\vert _{\xi _{Q}} \rightarrow 
\frac{16}{3\times 8}\frac{V_0}{\mP^4} 
\left[\left(\frac{\varphi _{\rm ini}}{\mP}\right)^{4}
-\left(\frac{\varphi }{\mP}\right)^{4}\right]\, ,
\end{eqnarray}
which is again the expression found in Ref.~\cite{ML}, specialized to
the case $n=2$. The previous conclusion, namely that this term
dominates the contribution coming from the inflaton noise, is not
modify in this particular case.

\subsection{Solution at Second Order}

We now turn to the calculation of the mean value of $\delta Q_2$ given
by the sum of five contributions as can be seen in
Eq.~(\ref{meanq2}). In principle, the calculations of these five terms
can be performed in the general case, where the classical quintessence
field in given either by Eq.~(\ref{qsolnneq2}) or
Eq.~(\ref{qsoln2}). However, for simplicity, we give only the
expressions corresponding to the case where $Q_\cl $ is frozen, \ie to
the limit $\Upsilon \rightarrow 0$. In this case, all the integrations
become trivial since they only involve integrals of power-law
functions. The first term, $\mean{\delta Q_2}\vert _{\delta \varphi
_2}$, originates from the term $\delta \varphi _2$ only. It is given
by
\begin{eqnarray}
\frac{\mean{\delta Q_2}}{\mP} \biggl \vert _{\delta \varphi _2} &=&
-\frac{\alpha }{12n}\frac{W_0}{\mP^4}\left(\frac{Q_{\rm
in}}{\mP}\right)^{-\alpha -1} \Biggl\{(n+2) \left[\left(\frac{\varphi
_\cl}{\mP}\right)^{2}-\left(\frac{\varphi _{\rm
in}}{\mP}\right)^{2}\right] +(n-2)\left(\frac{\varphi _{\rm
in}}{\mP}\right)^{4} \left[\left(\frac{\varphi
_\cl }{\mP}\right)^{-2}-\left(\frac{\varphi _{\rm
in}}{\mP}\right)^{-2}\right] 
\nonumber \\ 
& &+\frac{16}{n}\left(\frac{\varphi _{\rm in}}{\mP}\right)^{n/2+2}
\left[\left(\frac{\varphi _\cl}{\mP}\right)^{-n/2}-\left(\frac{\varphi
_{\rm in}}{\mP}\right)^{-n/2}\right]\Biggr\}\, , 
\end{eqnarray}
The main feature of the above term is the presence of the overall
factor $W_0/\mP^4\times \left(Q_\ini/\mP\right)^{-\alpha -1}$ (of
course, the powers of the inflaton field can also play a role but for
a crude order of magnitude estimate, one can ignore them). The second
term participating to the expression of $\mean{\delta Q_2}$ comes from
$\delta \varphi _1^2$. It is given by
\begin{eqnarray} 
\frac{\mean{\delta
Q_2}}{\mP} \biggl \vert _{\delta \varphi _1^2} =
\frac{\alpha(n+2)}{12n}\frac{W_0}{\mP^4}\left(\frac{Q_{\rm
in}}{\mP}\right)^{-\alpha -1} \Biggl\{\left[\left(\frac{\varphi
_\cl}{\mP}\right)^{2}-\left(\frac{\varphi _{\rm
in}}{\mP}\right)^{2}\right] +\left(\frac{\varphi _{\rm
in}}{\mP}\right)^{4} \left[\left(\frac{\varphi
_\cl}{\mP}\right)^{-2}-\left(\frac{\varphi _{\rm
in}}{\mP}\right)^{-2}\right]\Biggr\}\, .
\end{eqnarray}
We see that this term also scales as $W_0/\mP^4\times
\left(Q_\ini/\mP\right)^{-\alpha -1}$. Therefore, we expect the
previous contribution and this term to be of the same order of
magnitude. This can be checked explicitly in
Fig.~\ref{perturbative}. The third term comes from $\delta \varphi
_1\delta Q_1$ where it should be understood that, in $\delta Q_1$,
only the term proportional to the inflaton noise matters. One obtains
\begin{eqnarray}
& & \frac{\mean{\delta Q_2}}{\mP} \biggl \vert _{\delta \varphi _1\delta
Q_1} =-\frac{2\alpha ^2(\alpha
+1)}{3n^2}\frac{W_0}{\mP^4}\frac{W_0}{V_0} \left(\frac{Q_{\rm
in}}{\mP}\right)^{-2\alpha -3} \Biggl\{
\frac{1}{n}\left(\frac{\varphi _{\rm
in}}{\mP}\right)^{4} \left[\left(\frac{\varphi
_\cl}{\mP}\right)^{-n}-\left(\frac{\varphi _{\rm
in}}{\mP}\right)^{-n}\right]
\nonumber \\
& & -\frac{16}{n(8-n)}\left(\frac{\varphi _{\rm
in}}{\mP}\right)^{-n/2+4} \left[\left(\frac{\varphi
_\cl}{\mP}\right)^{-n/2}-\left(\frac{\varphi _{\rm
in}}{\mP}\right)^{-n/2}\right]
-\frac{n}{(4-n)(8-n)}\left[\left(\frac{\varphi
_\cl }{\mP}\right)^{-n+4}-\left(\frac{\varphi _{\rm
in}}{\mP}\right)^{-n+4}\right]\Biggr\}\, ,
\end{eqnarray}
In comparison with the two previous terms, we see that there is an
extra overall factor equal to $W_0/V_0\times
\left(Q_\ini/\mP\right)^{-\alpha -2}$. As a consequence, this term is
expected to be sub-dominant with respect to the two previous
contributions and, again, this can be checked explicitly in
Fig.~\ref{perturbative}. Of course, one could try to compensate the
smallness of $W_0/V_0$ by a large value of
$\left(Q_\ini/\mP\right)^{-\alpha -2}$, \ie by a large value of the
index $\alpha $ but this would lead to very artificial, hence
unphysical, models. Then, the fourth term, originating from the term
$\delta Q_1^2(\xi _Q)$, can be written as
\begin{eqnarray}
\frac{\mean{\delta Q_2}}{\mP} \biggl \vert _{\delta Q_1^2(\xi _Q)}
&=&\frac{8\alpha (\alpha +1)(\alpha +2)}{3n^2(n+2)}
\frac{V_0}{\mP^4}\frac{W_0}{V_0} \left(\frac{Q_{\rm
in}}{\mP}\right)^{-\alpha -3} \Biggl\{
\frac14\left[\left(\frac{\varphi
_\cl }{\mP}\right)^{4}-\left(\frac{\varphi _{\rm
in}}{\mP}\right)^{4}\right]
\nonumber \\
& & +\frac{1}{n-2}
\left(\frac{\varphi _{\rm
in}}{\mP}\right)^{n+2} \left[\left(\frac{\varphi
_\cl }{\mP}\right)^{2-n}-\left(\frac{\varphi _{\rm
in}}{\mP}\right)^{2-n}\right]\Biggr\}\, .
\end{eqnarray}
If we compare this term with the two first contributions, we see that
there is an extra overall factor equal to
$\left(Q_\ini/\mP\right)^{-2}$. Since $Q_\ini/\mP $ is small one
expects the previous contribution to be dominant. In
Fig.~\ref{perturbative}, this contribution is represented by the solid
black line which is indeed the most important one. Finally, the last
term coming, from $\delta Q_1^2(\xi )$, reads
\begin{eqnarray}
\frac{\mean{\delta Q_2}}{\mP} \biggl \vert _{\delta
Q_1^2(\xi)} &=& \frac{2\alpha ^3(\alpha +1)(\alpha +2)}{3n^3}
\frac{W_0}{\mP^4}\frac{W_0^2}{V_0^2} \left(\frac{Q_{\rm
in}}{\mP}\right)^{-3\alpha -5} \Biggl\{ \frac{n}{2(3-n)(4-n)(8-n)}
\left[\left(\frac{\varphi _\cl}{\mP}\right)^{6-2n}-\left(\frac{\varphi
_{\rm in}}{\mP}\right)^{6-2n}\right] 
\nonumber \\ & &
+\frac{32}{n(4-3n)(8-n)} \left(\frac{\varphi _{\rm
in}}{\mP}\right)^{4-n/2} \left[\left(\frac{\varphi
_\cl}{\mP}\right)^{2-3n/2}-\left(\frac{\varphi _{\rm
in}}{\mP}\right)^{2-3n/2}\right] 
\nonumber \\ & & 
+\frac{1}{2n(n-1)}
\left(\frac{\varphi _{\rm in}}{\mP}\right)^{4}
\left[\left(\frac{\varphi _\cl}{\mP}\right)^{2-2n}
-\left(\frac{\varphi
_{\rm in}}{\mP}\right)^{2-2n}\right] 
\nonumber \\ & &
+\frac{4}{n(n-2)(4-n)} \left(\frac{\varphi _{\rm
in}}{\mP}\right)^{4-n} \left[\left(\frac{\varphi
_\cl}{\mP}\right)^{2-n}-\left(\frac{\varphi _{\rm
in}}{\mP}\right)^{2-n}\right]\Biggr\}\, .
\end{eqnarray}
\end{widetext}
This term is proportional to $(W_0/V_0)^2$ which is a tiny
factor. Therefore, the above contribution is expected to be the
smallest contribution to $\mean{\delta Q_2}$ and, in fact, to be
totally negligible. This is confirmed in Fig.~\ref{perturbative}.

\end{document}